\documentclass[11pt,a4paper]{article}
\usepackage{jcappub}
\usepackage{lscape}
\usepackage{multirow}

\newcommand{\Lor}[2]{\Lambda^{#1}_{\;\;#2}}

\newcommand{\rEF}{r^*}
\newcommand{\Rcal}{{\cal R}}

\newcommand{\qsol}{q}
\newcommand{\qA}{q_{A}}

\newcommand{\charge}{q_{BH}}
\newcommand{\rcharge}{\ensuremath{\sigma}}
\newcommand{\sA}{\ensuremath{s_A}}

\newcommand{\sE}{\ensuremath{s_E}}

\newcommand{\su}{\ensuremath{s_{u_0}}}

\newcommand{\schi}{\ensuremath{s_{\chi_0}}}
\newcommand{\Sign}[1]{\ensuremath{{\rm Sign}(#1)}}
\newcommand{\Rp}{\ensuremath{r_{+}}}
\newcommand{\Rm}{\ensuremath{r_{-}}}
\newcommand{\Rpm}{\ensuremath{r_{\pm}}}

\newcommand{\grad}{\ensuremath{\vec{\nabla}}}
\newcommand{\covec}[1]{\ensuremath{ \underline{#1} }}

\newcommand{\Ab}{{\bar{A}}}

\newcommand{\lambdas}{\lambda_s}

\newcommand{\mt}{\tilde{m}} 
\newcommand{\nt}{\tilde{n}}

\newcommand{\Kcal}{{\cal K}}

\newcommand{\nhat}{\hat{n}}

\newcommand{\phih}{\hat{\phi}}
\newcommand{\phit}{\tilde{\phi}}

\newcommand{\Ocal}{{\cal O}}

\newcommand{\Fcal}{{\cal F}}

\newcommand{\Qcal}{{\cal Q}}
\newcommand{\Qcalh}{ \hat{{\cal Q}}}

\newcommand{\Ycal}{{\cal Y}}

\newcommand{\Jcal}{{\cal J}}

\newcommand{\Gt}{\ensuremath{\tilde{G}}}
\newcommand{\GN}{\ensuremath{G_{{\rm N}}}}
\newcommand{\KB}{\ensuremath{K_{{\rm B}}}}

\newcommand{\sEF}{s_{EF}}
\begin{document}

\title{Stealth black holes in Aether Scalar Tensor theory}

\author[a,b]{Constantinos Skordis}

\author[a,c]{David M. J. Vokrouhlický}%
\affiliation[a]{
  CEICO, Institute of Physics of the Czech Academy of Sciences, Na Slovance 1999/2, 182 00, Prague, Czechia
}
\affiliation[b]{
 Department of Physics, University of Oxford, Denys Wilkinson Building, Keble Road, Oxford OX1 3RH, United Kingdom
}
\affiliation[c]{
Institute of Theoretical Physics, Faculty of Mathematics and Physics, Charles University, V Holešovičkách 747/2, 180 00 Prague 8, Czechia
}
\date{\today}

\emailAdd{skordis@fzu.cz}
\emailAdd{vokrouhlicky@fzu.cz}

\abstract{
The Aether Scalar Tensor (AeST) theory is an extension of general relativity (GR) successful at reproducing galactic rotational curves, gravitational lensing, 
 linear large scale structure and cosmic microwave background power spectrum observations. 
We solve the most general static spherically symmetric vacuum equations in the strong-field regime of AeST 
and find two classes of stealth black hole solutions -- those with exact GR geometries --  containing non-trivial secondary hair. 
In particular, one of these can be continuously joined to the cosmological solution of AeST. 
We also derive a non-black hole solution with zero spatial component in the vector field.
This result proves the existence of mathematically and observationally consistent candidates for black holes in AeST, 
and creates a basis for testing the theory in the strong-field regime.}


                             
\maketitle


\section{\label{sec:introduction}Introduction}
General Relativity (GR) has been widely successful in explaining and predicting many gravitational phenomena. It has so far passed several precise 
observational tests~\cite{will2014confrontation} ranging from table-top experiments~\cite{adelberger2003tests}, our solar system~\cite{ni2016solar} 
and other astrophysical systems~\cite{berti2015testing}, to the detection of gravitational waves from mergers of compact objects~\cite{PhysRevLett.119.161101}. 
Nevertheless, there are reasons for us to consider extensions to GR~\cite{Clifton:2011jh}. The most well-motivated 
extension is that of added higher curvature terms which come as quantum corrections to GR~\cite{Donoghue:1994dn,Burgess:2003jk,Woodard:2009ns} and
 treated in the language of effective field theory.
Another possibility which has been in vogue in the last twenty years or so, is to extend GR with new gravitational degrees of freedom playing the role of
dark energy, in an attempt to alleviate 
some  of the issues associated with the cosmological constant~\cite{Clifton:2011jh,Joyce:2014kja,Joyce:2016vqv}.
The less investigated case, which is more the concern of this article, is to do with the missing mass commonly attributed to dark matter.

GR is the cornerstone of our current cosmological model -- the cold dark matter model with a cosmological constant ($\Lambda$CDM). 
These two ingredients are necessary, within GR, to explain the observed mismatch between the observed dynamics of luminous matter and its gravitational influence. 
Of these, dark matter is commonly believed to be in the form of particles which are not part of the standard model of particle physics, however, 
such particles have so far been elusive to particle searches. As such, the possibility that an extension of GR may explain this mismatch without
the presence of dark matter, remains open.

Perhaps the most widely investigated scenario is that of Modified Newtonian Dynamics (MOND)~\cite{Milgrom:1983ca,Milgrom:1983pn,Milgrom:1983zz} which was prescribed as a phenomenological model introducing an acceleration scale $a_0=1.2\cdot 10^{-10}\text{m/s}^2$ under which the law of Newtonian gravity is altered; see~\cite{famaey2012modified} for a review. MOND was quickly cast as an extension of Newtonian gravity~\cite{1984ApJ...286....7B} which was also shown to emerge in the weak-field limit of a scalar-tensor extension of GR in the same work. That early model was shown to be incompatible with observations of gravitational lensing from bounded baryonic structures and the extensions that followed~\cite{Bekenstein:1988zy} did not improve on this. The use of a scalar-field based disformal relation between two metrics was proposed~\cite{Bekenstein:1992pj} but without overcoming the problem with gravitational lensing~\cite{Bekenstein:1993fs}. The latter was resolved by Sanders using a unit-timelike vector field in the disformal relation~\cite{Sanders:1996wk} and the resulting model was later developed by Bekenstein into a covariant theory, the Tensor-Vector-Scalar (TeVeS) theory~\cite{PhysRevD.70.083509}. The formulation of TeVeS initiated a stronger interest in the possibility of extensions of GR as alternatives to the particle dark matter hypothesis, as it became possible to study other phenomena resulting from such extensions, particularly cosmology~\cite{Skordis:2005xk,Dodelson:2006zt,Bourliot:2006ig}, black holes~\cite{giannios2005spherically,sagi2008black,Lasky:2010bd}, neutron stars~\cite{lasky2008structure}, solar system tests~\cite{Sagi:2009kd,Sagi:2010ei}, Hamiltonian analysis~\cite{Chaichian:2014dfa}, and gravitational waves~\cite{Sagi:2010ei,Gong:2018cgj,Skordis:2019fxt}. Further extensions or generalizations of TeVeS followed~\cite{Sanders:2005vd,Skordis:2008pq,babichev2011improving,Zlosnik:2017xpr}, as well as other GR extensions not related to TeVeS but with similar purpose~\cite{Zlosnik:2006zu,PhysRevD.80.123536,Zuntz:2010jp,Blanchet:2011wv,Sanders:2011wa,deffayet2011nonlocal,Mendoza:2012hu,Khoury:2014tka,deffayet2014field,Verlinde:2016toy,Burrage:2018zuj,PhysRevD.100.084039,DAmbrosio:2020nev,deffayet2024price,Blanchet:2024mvy}. See~\cite{bruneton2007field} for an early review of field-theoretical formulations of MOND~\footnote{We note that extended dark matter models~\cite{Blanchet:2006yt,Blanchet:2009zu,Berezhiani:2015pia,Berezhiani:2015bqa,Kaplinghat:2015aga,Kamada:2016euw} have also been proposed to accommodate for MOND behaviour, keeping GR as the description of gravity. The possibility of a bi-metric theory (and thus GR extension) with dark matter, as a way of leading to MOND behaviour was also investigated~\cite{Blanchet:2015sra}.}.

The usual obstacle that GR extensions have to face is to reproduce the success of large scale cosmology and $\Lambda$CDM phenomenology in the absence of a dark matter particle. Moreover,
extending GR with the introduction of new gravitational degrees of freedom quite often leads to a propagation speed for the tensor-mode gravitational wave different than the speed of light. However, this has been severely constrained~\cite{LIGOScientific:2017vwq,Savchenko:2017ffs,Goldstein:2017mmi}. In cases where classic cosmological observables have been computed -- the matter power spectrum and the cosmic microwave background angular power spectra -- none of the above theories have been shown to fit all of the cosmological data while preserving MOND phenomenology in galaxies. 
The first model where this was shown to be possible, using the same fields both for cosmology and for MOND, is the Aether Scalar Tensor theory (AeST)~\cite{2021PhRvL.127p1302S}.

The AeST theory uses the same fields as the old TeVeS proposal, the metric, a scalar and a unit-timelike vector field, however, it removes the necessity of a disformal transformation in its formulation. It was conceived to lead to MOND phenomenology in its weak field limit and correct gravitational lensing as in TeVeS theory. However, unlike TeVeS, it can fit the large scale cosmological spectra similarly as $\Lambda$CDM, while the tensor modes of the theory propagate at the speed of light. Several articles study this theory in a variety of observational or theoretical setups~\cite{2022PhRvD.106j4041S,Kashfi:2022dyb,Mistele:2021qvz, PhysRevD.107.044062, 2023A&A...676A.100M,2023GReGr..55...23B,llinares2023extension,Verwayen:2023sds,Durakovic:2023out,Bataki:2023uuy,Mistele:2023fwd,Rosa:2023qun,reyes2024neutronstarsaetherscalartensor,babichev2015charged}. 

An important feature of every viable theory is the presence of black hole solutions, or at least compact objects, near those described by GR, as there is strong observational evidence for their existence \citep{PhysRevLett.119.161101, collaboration2019first}. These objects and their departures from pure GR can serve as an important probe for distinguishing the presence of additional fields and various extensions of GR \cite{yagi2016black}. These departures can be both in the geometry of such objects and in the presence of hair \cite{coleman1992quantum}, meaning the presence of stable additional fields with non-trivial information content. Hair is further classified into \emph{primary} and \emph{secondary}, 
where primary add a new charge characterizing the solutions, whereas secondary do not. Some theories can contain stealth black holes - black holes whose geometries are solutions to the vacuum Einstein equations - but for which the additional fields have non-trivial solutions (hair)~\cite{ayon2006stealth}. The exact definition slightly varies in the literature and we choose a more liberal approach~\footnote{Some authors define stealth BHs as those with a Ricci flat or an Einstein metric~\cite{bakopoulos2024black}, while others extend the nomenclature to Einstein-Maxwell theory and classify solutions where the metric is of Reissner-Nordstrom type as stealth~\cite{fan2018black,baake2024endowing}.} of defining stealth black holes as BHs with geometries of solutions to GR coupled with standard matter fields as in~\cite{chagoya2018stealth}. This makes stealth BHs geometrically indistinguishable from GR black holes, and one must consider other signatures such as quasinormal modes~\cite{de2019perturbations} and other perturbative signatures~\cite{de2023approximately}, or thermodynamics~\cite{bakopoulos2024stealth,Erices:2024iah} as relevant distinguishing probes.
Stealth black holes have been found in theories with similar components as AeST - shift-symmetric scalar fields \cite{minamitsuji2018stealth,minamitsuji2019black, bernardo2020stealth} and vector-tensor theories \cite{heisenberg2017hairy, chagoya2018stealth, minamitsuji2016solutions}. 

Compact spherically symmetric objects have been studied in the context of the old TeVeS theory. Giannios found the first TeVeS static spherically symmetric solutions and a subset of these describe stealth BH solutions~\cite{ giannios2005spherically} which, however, led to superluminal propagation of scalar perturbations. Bekenstein and Sagi found stealth Reissner-Nordstrom (RN) type solutions~\cite{sagi2008black} where the RN charge was provided by the vector field present in TeVeS, which contained the Giannios solution but also a different branch which did not have the superluminality problem. Neutron star solutions were studied in~\cite{lasky2008structure} while quasinormal modes were studied in~\cite{Lasky:2010bd}. A comprehensive summary of these solutions can be found in~\cite{Skordis:2009bf}. 

Given the few compact object solutions already found in TeVeS, some of which are of the stealth BH type, it is relevant to investigate the existence of compact object solutions in AeST. A first investigation of BH solutions in AeST can be found in~\cite{2023GReGr..55...23B}. There, the authors present a general construction approach drawn from Horndeski models for finding solutions without specifying the full Lagrangian in this theory. Using their procedure, they found 
two concrete solutions with very specific parameters which, however, lie outside the regime of stability of the theory~\cite{2022PhRvD.106j4041S}: a non-black hole solution and a very specific case of a hairy Schwarzschild black hole. Specific cases of neutron star solutions was concluded in~\cite{reyes2024neutronstarsaetherscalartensor} alongside the derivation of a Tolman–Oppenheimer–Volkoff-like equation for this theory.

In this article we consider the general strong-field regime of the AeST theory that fully fixes the free function present in its Lagrangian and closes the system of the Einstein, scalar and vector field equations. We solve this system in the most general spherically symmetric static vacuum configuration leading to all possible classes of solutions under these assumptions. The paper is organized as follows: section \ref{sec:theory} presents the AeST theory and its vacuum equations in the strong field regime, alongside details about our ansatz. 
In section \ref{sec:solutions_general}, we solve these equations in a general setup
 under the  assumption that the vector field has a spatial component which is non-zero but determined by the field equations.
We follow in section  \ref{sec:solutions_special} with the case where the spatial component of the vector field is zero everywhere, filling the gap in the derivation of the solutions
of section  \ref{sec:solutions_general}.
In section \ref{sec:discussion} we make some more comments on the consistency of found solutions, including their weak-field limits. Section \ref{sec:conclusion} concludes the findings of this paper and discusses further considerations. We use Greek letters to denote spacetime indices taking values in $0\ldots 3$, 
capital Latin letters to denote angular indices taking values in $2\ldots 3$, and use a $-+++$ signature and the curvature conventions of Wald~\cite{wald:1984}. We also define the usual (anti)symmetrization conventions $2X_{(\mu\nu)} \equiv X_{\mu\nu} + X_{\nu\mu}  $ and $2X_{[\mu\nu]} \equiv X_{\mu\nu} -X_{\nu\mu}  $.

\section{\label{sec:theory}Theory and field equations}
The AeST theory contains a metric $g_{\mu\nu}$, a scalar field $\phi$, and a unit time-like vector field $A^\mu$. The action of the theory is
\begin{equation}
\begin{aligned}
S_g=\int d^4 x \frac{\sqrt{-g}}{16 \pi \tilde{G}}[\mathcal{R} & -\frac{K_B}{2} F^{\mu \nu} F_{\mu \nu}+2\left(2-K_B\right) J^\mu \nabla_\mu \phi \\
& \left.-\left(2-K_B\right) \Ycal-\Fcal(\Ycal, \Qcal)-\lambda\left(A^\mu A_\mu+1\right)\right],
\end{aligned}
\end{equation}
where $\nabla_\mu$ is the covariant derivative, $\lambda$ the Lagrange multiplier normalizing the vector field, $\Gt$ the bare gravitational strength, $\KB$ a constant parameter of the theory similar to TeVeS theory, $\Rcal$ is the Ricci scalar and $F_{\mu \nu}\equiv\nabla_\mu A_\nu-\nabla_\nu A_\mu$. We define the metric on orthogonal subspaces to $A_{\mu}$ as $q_{\mu\nu}=g_{\mu\nu}+A_\mu A_{\nu}$ and denote the associated covariant derivative as $D_{\mu} \equiv q_\mu^{\;\;\nu} \nabla_\nu$. Scalars in the Lagrangian are defined as $\Qcal\equiv A^\mu\nabla_\mu\phi$ and $ \Ycal \equiv D^\mu\phi D_\mu\phi$ and the vector quantity $ J_\nu\equiv A^\mu\nabla_\mu A_\nu$. The function $\Fcal$ is generally left as a free function in the Lagrangian, however, it must obey specific constraints in order to have correct cosmological evolution and Newtonian and MOND limits. Note that the Lagrangian does not explicitly  depend on $\phi$; the theory is shift symmetric in this scalar field. 

\subsection{Strong-field regime with correct cosmology} 
We set the theory to be in the strong-field regime by letting the free function of the theory $\Fcal$ take on the form
\begin{align}
\Fcal =  \left(2-\KB\right) \lambdas \Ycal - 2 \Kcal_2 \left(\Qcal - \Qcal_0\right)^2+\ldots,
\label{Fcal:strong}
\end{align}
where $\lambdas$ and $\Kcal_2$ are constants. The first term in \eqref{Fcal:strong} is necessary for having a consistent strong-field regime while the second term
is the lowest order in an expansion in $\Qcal - \Qcal_0$ which leads to Friedman-Lema\^{i}tre-Robertson-Walker (FLRW) evolution for the energy density of the scalar  to be that of dust as is necessary for fitting CMB observations~\cite{2021PhRvL.127p1302S}. We do not include MOND terms of the form $\Ycal^{3/2}$, since we are interested in compact solutions with scales much smaller than the MOND radius. It was shown in~\cite{2022PhRvD.106j4041S} that stability of linear perturbations on Minkowski spacetime requires the conditions
$0 < \KB < 2$, $\Kcal_2 >0$ and $\lambdas\ge0$ which we assume in the remaining of the article. With these in mind
we define $\mu^2 \equiv 2\Kcal_2 \Qcal_0^2/(2-\KB)$ which is related to the scale of validity of Minkowski spacetime~\cite{2022PhRvD.106j4041S} 
as well as the quantities 
\begin{equation}
\nt =  \frac{2+\KB\lambdas}{2(1+\lambdas)}
\label{def_nt}
\end{equation}
and 
\begin{equation}
\mt =  \sqrt{\frac{2-\KB}{2+\KB\lambdas}},
\label{def_mt}
\end{equation}
obeying $0<\nt\leq1$ and $0<\mt<1$ which appear frequently in what follows.

Setting $\Fcal$ to \eqref{Fcal:strong} and varying the action with respect to the scalar field yields 
\begin{align}
\nabla_\mu S^\mu =& 0,
\quad
 & S^\mu\equiv& ( 2 - \KB)(1+\lambdas) D^\mu \phi - \left(2 - \KB\right)J^\mu -  2\Kcal_2 \left(\Qcal - \Qcal_0\right)    A^\mu,
\label{scalar_eq}
\end{align}
where $S^\mu$ is the Noether current of the shift symmetry, 
while the vector field equation is found as
\begin{align}
&
 \frac{\KB}{2-\KB}  q_\mu^\rho   \nabla_\nu F^\nu_{\;\;\rho} +2\nabla^\nu \phi  D_\mu A_\nu
- \left[(1+\lambdas) \Qcal + J - \frac{2 \Kcal_2\left(\Qcal - \Qcal_0\right) }{2-\KB}  \right] D_\mu \phi 
-  D_\mu \Qcal
 =0,
\label{vector_eq}
\end{align}
where we have defined $J \equiv \nabla_\mu A^\mu$.
Finally, the vacuum Einstein equations take the form
\begin{align}
 G_{\mu\nu} =&  (2-\KB) \left[ (1+\lambdas)   D_\mu \phi D_\nu \phi 
 - 2   \nabla_{(\mu} \phi J_{\nu)} 
 - 2 \nabla_\rho \phi \left( A_{(\mu} \nabla_{\nu)} A^\rho - A_{(\mu} \nabla^\rho  A_{\nu)} \right)
 \right]
\nonumber
\\
&
 - \KB F^\rho_{\;\;(\mu} F_{\nu)\rho} 
+ \frac{1}{2} C^{(g)}  g_{\mu\nu} 
+ C^{(A)} A_\mu A_\nu
- 4 \Kcal_2 \left(\Qcal - \Qcal_0\right)   D_{(\mu} \phi A_{\nu)} ,
\label{Einstein_eq}
\end{align}
where we have defined
\begin{align}
C^{(g)} \equiv& - \frac{\KB}{2} F_{\alpha\beta}  F^{\alpha\beta} 
+  \left(2 - \KB\right) \left[ 2J^\rho D_\rho \phi -  (1+\lambdas) \Ycal \right] + 2 \Kcal_2 \left(\Qcal - \Qcal_0\right)^2,
\label{def_C_g}
\\
C^{(A)} \equiv& \left(2 - \KB\right) \left( q^\lambda_{\;\;\rho}\nabla_\lambda\nabla^\rho \phi + J \Qcal- D^\rho \phi J_\rho \right)
-\KB A^\rho  \nabla_\lambda F^\lambda_{\;\;\rho}
+ 2 \Kcal_2 \left(\Qcal - \Qcal_0\right) \Qcal .  
\label{def_C_A}
\end{align}
to collect together these terms. Varying with respect to the  Lagrange multiplier keeps the vector field normalized, that is, $A_\mu A^\mu + 1=0$.

\subsection{Spherically symmetric static solutions}
We impose spherical symmetry and staticity for the metric using the Schwarzschild-like coordinates $t$ and $r$ where the metric takes the form
\begin{align}
 ds^2 = - e^{2\Phi} dt^2  +  e^{2\Psi}  dr^2 + r^2 d\Omega^2
\label{eq_SSS_metric}
\end{align}
where $\Phi = \Phi(r)$, $\Psi = \Psi(r)$ and $d\Omega^2$ is the metric of a 2-sphere.  We
denote differentiation with respect to the radial coordinate using the $\prime$ symbol, i.e. $A' \equiv dA/dr$.

The notion of staticity for the fields is slightly more nuanced for the fields than the metric. Since the Lagrangian of the theory is only proportional to gradients of the scalar field and the theory is shift-symmetric, a linear term $t$ in the scalar field is still a static configuration, as $t$ would not appear anywhere in the field equations or the action. This means that foliations of the metric with constant $t$ and surfaces of constant scalar field do not necessarily coincide, but the gradient of the scalar field is conserved on curves of constant $r$. Putting it differently, the scalar field can sometimes be used as a time coordinate, which is called the unitary gauge. With these considerations we have that 
\begin{align}
\phi = \Qcal_0 R_1(r) t + R_2(r).
\end{align}
This is the most general static spherically symmetric scalar field. Terms mixing $t$ and $r$ would not correspond to a static configuration, and dependence on $t$ must be linear. However, this is
not sufficient as with this ansatz, important scalars of the Lagrangian, such as $\Qcal$, would be time-dependent. Therefore, our final ansatz reduces to
\begin{equation}
    \phi=\Qcal_0 \left( \qsol  t + R \right),
    \label{phi_ansatz}
\end{equation}
where $R = R(r)$ only and $\qsol$ is a constant integer which we use to distinguish \emph{solution} classes and only takes the values $\qsol=0$ or $\qsol = 1$. 
We briefly discuss the possibility $\qsol = -1$ below.
The ansatz \eqref{phi_ansatz} is commonly used in other shift-symmetric theories, for further justification and references see~\cite{babichev2014dressing, kobayashi2014exact, charmousis2015self}. 

In order for the scalar to be continuously matched to the cosmological FLRW solution, its gradient must be timelike. In the converse case, that is, if its gradient were spacelike, there would be a surface of discontinuity in the gradient, or a type of acoustic horizon, surrounding bounded structures such as black holes or stars, if they are embedded in an FLRW cosmological background. Such a situation does not seem physical.
Thus, the only physical solutions are those with $\qsol=1$, however, we find also the solution class
$\qsol=0$ for completeness, as it could find use elsewhere.

We now turn to the vector field. Its components in the angular directions vanish as a result of the spherical symmetry, that is $A_2 = A_3 = 0$. Moreover,
the unit-timelike constraint relates its $A_0$ component with its $A_1$ component and so only one independent function $A(r)$ remains which we choose to be the component $A_1 = A(r)$. Solving the unit-timelike constraint leads to $A_0 = - e^\Phi \chi$ where  
\begin{equation}
\chi \equiv  \pm \sqrt{1+A^2e^{-2\Psi}}.
\label{chi_def}
\end{equation}
In what follows, we will consider $\chi$ to carry the sign due to taking the square root above.~\footnote{It might seem that variables such as $\chi$ and $e^{\Phi}$ are not always well-defined, for instance, under horizons. However, components of the fields of the theory are always well-defined. As an example, 
we can see that in $A_0=\pm\sqrt{\chi^2 e^{2\Phi}}$, taking the square root of either variable would be pathological, as both are negative, but taking the root 
of the combination $\chi^2 e^{2\Phi}$ yields a consistent result.} 

\subsubsection{Discrete symmetry}
\label{sec:discrete_symmetry}
Setting the function $\Fcal$ aside, the action has the discrete symmetry  $A_\mu \rightarrow - A_\mu$. 
Correct cosmological behaviour can be expressed with $\Fcal$ having the expansion
$\Fcal = - 2 \Kcal_2 \left(\Qcal - \Qcal_0\right)^2 + \ldots$ where this symmetry seems to be generally broken,
although functional forms of $\Fcal$ where the symmetry is intact, and correct cosmological behaviour still emerges 
can be constructed, for instance, $\Fcal \propto \left(\Qcal^2 - \Qcal_0^2\right)^2$. Nevertheless, as we will discuss later, in the small scale limit $\mu\rightarrow 0$ the symmetry holds and we use it to simplify the classification of solutions.

\subsubsection{Asymptotic frame freedom\label{sec:asymptotics}}
A general Lorentz transformation $\Lor{\mu}{\mu'}$ preserves the Minkowski metric $\eta_{\mu\nu}$. Starting from the spherical system $\{ t, r,\theta,\phi\}$ with a vector field 
in the form of $A_\mu = ( - (1 + \Ab^2)^{1/2}, \Ab,0,0)$ with $\Ab$ a constant, we can then choose a Cartesian system aligned with the $A_i$ axis 
so that the above form may be thought to hold in that system. We can then  Lorentz-boost away the $A_i$ component and bring $A_\mu$ into
the form  $A_\mu = (-1, 0 ,0 ,0 )$.
A  second vector field, for instance, $\nabla_{\mu}\phi$,  need not  in general be in the same frame as $A_\mu$.  
If we look at asymptotically flat solutions, we have the freedom perform Lorentz transformations to the two fields $A_\mu$ and $\nabla_\mu\phi$ at $r\rightarrow\infty$ 
without affecting the asymptotic metric.  Specifically, in spherical symmetry there is only one spatial axis available which is the
axis aligned with $r$, thus, as expected (and as we show below), the two asymptotic frames coincide, leading to simplifications of the solution. 
The reader is referred to appendix-\ref{sec:appendix_lorentz} for more details.

\subsection{The field equations}
We now present the field equations adapted to our ansatz. To aid the reader, we present some of the  calculational steps in appendix \ref{sec:appendix_calc}.

\subsubsection{The scalar and vector field equations}
Rather than computing second derivatives of $A$, it is more useful to define the quantity $E \equiv e^{-\Phi} F_{01}$ which evaluates to
\begin{equation}
    E = \chi'+\chi \Phi'.
    \label{eq_def_E}
\end{equation}
We also define the frequently used variable $C^{(\Qcal)}$ as
\begin{align}
C^{(\Qcal)} \equiv   \frac{\Qcal}{\Qcal_0} - 1   =    e^{-\Phi} \chi \qsol +e^{-2\Psi} A R'- 1 ,
\label{def_C_mu}
\end{align}
where the last equality has been evaluated using our static, spherical symmetric ansatz \eqref{eq_SSS_metric} and \eqref{phi_ansatz}.
Since the theory is shift-symmetric, the scalar equation \eqref{scalar_eq}
 is of the form $\nabla_\mu S^\mu=0$ for a current $S^\mu$ and can be integrated to obtain a Noether charge corresponding  to the shift symmetry.
In the SSS case we find that upon this integration the scalar equation gives
\begin{align}
    \Qcal_0\chi E - (1+\lambdas)\Qcal_0^2\chi\left[ e^{-\Phi}  A \qsol +\chi R'\right] + \mu^2 A  C^{(\Qcal)} = \frac{\phi_0  e^{-\Phi+\Psi} }{r^2},
\label{reduced_scalar_SSS}
\end{align}
 where $\phi_0$ is an integration constant. 

Due to the unit-norm constraint on the vector field, the $\mu=0$ and $\mu=1$ components of the vector equation \eqref{vector_eq} are linear combinations of each other and reduce to
\begin{align}
&
\frac{\KB}{2-\KB}  e^{\Phi-2\Psi}    A  \left[ E' + \left( - \Psi' + \frac{2}{r} \right) E \right]
-  \Qcal_0 \qsol  \Phi'
 - 2 \qsol \Qcal_0 e^{-2\Psi} A^2\left(  \Phi'  -  \frac{1}{r}  \right)   
\nonumber
\\
&
+ \left[ (1+\lambdas)  \Qcal_0^2 \left( e^{-\Phi}  \chi \qsol + e^{-2\Psi} A R' \right)  - \mu^2  C^{(\Qcal)}  \right] \left(A \qsol + e^{\Phi} \chi R' \right)
\nonumber
\\
&
+ \Qcal_0 e^{\Phi-2\Psi}  \chi A  \left\{ R'' +  \left[-(\Phi'+\Psi')+\frac{2}{r}\right]R' \right\}
 =0,
\label{reduced_vector_SSS}
\end{align}
which is equal to the $\mu=1$ component of \eqref{vector_eq} multiplied by $- \frac{e^{\Phi}}{  (2-\KB)\chi}$. The angular components are identically zero, as expected.

\subsubsection{The Einstein equations}

The $G_{01}$ component of the Einstein equation \eqref{Einstein_eq} divided by $2-\KB$, leads to the relation
\begin{align}
 0 =& 
- \frac{\KB }{2-\KB} e^{\Phi-2\Psi}  \chi^2 A\left[ E' + \left( - \Psi' + \frac{2}{r} \right) E \right]
-  \qsol \Qcal_0   \chi   \left[   E -  \chi \Phi' - 2 e^{-2\Psi} A^2 \chi \left(   \Phi'  -  \frac{1}{r} \right) \right] 
\nonumber
\\
& 
- (1+\lambdas) \Qcal_0^2   e^{\Phi-2\Psi} A  \chi  \left(e^{-\Phi} A \qsol + \chi R'\right)^2  
-  \Qcal_0  e^{\Phi-2\Psi} \chi^3 A  \left[ R'' + \left( -    \Psi' -   \Phi' +  \frac{2}{r} \right)  R' \right]
\nonumber
\\
&
 + \mu^2  C^{(\Qcal)}   \left(  e^{-2\Psi} A^3  \qsol +  e^{\Phi} \chi^3 R' \right).
 \label{G_01_SSS}
\end{align}
We add the above equation \eqref{G_01_SSS} to  $\chi^2$ times \eqref{reduced_vector_SSS} and then add \eqref{reduced_scalar_SSS} to find the constraint on the scalar charge:
\begin{align}
  \frac{\phi_0  e^{-\Phi+\Psi} }{r^2}
  =& 
\left(1-\qsol \right) \left[
  \Qcal_0   \chi      E 
- (1+\lambdas)  \Qcal_0^2    \chi \left(e^{-\Phi}  A \qsol +  \chi R'\right) 
 + \mu^2  A C^{(\Qcal)}
\right].
\label{Phi_Psi_mu}
\end{align}
Thus, in the case $\qsol=1$ we find $\phi_0  = 0$,  
while in the case $\qsol=0$, \eqref{reduced_scalar_SSS} holds without any constraint on $\phi_0$. These two cases are combined in the constraint
\begin{align}
\qsol \phi_0  = 0.
\label{scalar_charge}
\end{align}

We now turn to the other Einstein equations.
The $G_{00}$ component of the Einstein equation \eqref{Einstein_eq} multiplied by $e^{2(\Psi-\Phi)}$ gives
\begin{align}
  \frac{2}{r}   \Psi' 
  +\frac{ e^{2\Psi}  - 1}{r^2}  
  =& 
 e^{2\Psi} \chi^2  C^{(A)} 
- \frac{1}{2} e^{2\Psi}  C^{(g)}   
+ \KB E^2  
+ (2-\KB) \left(  e^{-\Phi} A \qsol +  \chi   R' \right)\bigg[
 2 \Qcal_0 E 
\nonumber
\\
&
+ (1+\lambdas)  \Qcal_0^2  e^{-2\Psi} A^2  \left(e^{-\Phi} A  \qsol + \chi R'\right)
 - 2 \mu^2  A \chi  C^{(\Qcal)}
\bigg].
\label{reduced_Ein_00}
\end{align}
The $G_{11}$ component of the Einstein equation \eqref{Einstein_eq} reduces to
\begin{align}
 \! \! \! 
 \frac{2}{r}   \Phi'   -\frac{e^{2\Psi}-1}{r^2}   
  =& 
   A^2  C^{(A)}   + \frac{1}{2} e^{2\Psi} C^{(g)} 
  -\KB E^2   
+ (2-\KB)   \left( e^{-\Phi} A \qsol + \chi R' \right) \bigg[  
 - 2 \Qcal_0  E  
\nonumber
\\
&
+ (1+\lambdas) \Qcal_0^2 \chi^2  \left( e^{-\Phi} A \qsol + \chi R' \right)
 - 2 \mu^2  A \chi C^{(\Qcal)}
\bigg],
\label{reduced_Ein_11}
\end{align}
and finally, the $G_{AB}$ Einstein equation \eqref{Einstein_eq} multiplied by $e^{2\Psi}/r^2$ leads to
\begin{align}
\! 
\Phi'' + \left(\Phi' - \Psi' \right) \left(\Phi' +  \frac{1}{r} \right) 
 =&  
  (2-\KB) \left(e^{-\Phi} A \qsol  + \chi R'\right)  \left[  \Qcal_0  E - \frac{1}{2} (1+\lambdas) \Qcal_0^2  \left(e^{-\Phi} A  \qsol + \chi R'\right) \right]
\nonumber
\\
&
+ \frac{\KB  }{2}  E^2
+ \frac{1}{2} (2-\KB) \mu^2 e^{2\Psi} \left( C^{(\Qcal)} \right)^2.
\label{reduced_Ein_AB}
\end{align}
To summarize, we have reduced the relevant equations to \eqref{eq_def_E}, \eqref{reduced_scalar_SSS}, \eqref{scalar_charge}, \eqref{reduced_Ein_00}, \eqref{reduced_Ein_11} and \eqref{reduced_Ein_AB}, all of which hold for any $A$, even if it is zero everywhere. Below we find solutions to these equations. As we show, we do not need anymore \eqref{reduced_vector_SSS} and \eqref{G_01_SSS} which can be used, however, to test consistency of the final solutions.

\section{General solutions with $A\ne 0$}
\label{sec:solutions_general}

We now consider the main cases of this article. We assume that $A\ne 0$ except at a set of measure zero, so that we can freely divide by $A$. We proceed first by
adding \eqref{reduced_Ein_00} and \eqref{reduced_Ein_11} and using \eqref{reduced_vector_SSS} to eliminate $E'$
and \eqref{reduced_scalar_SSS} to eliminate $E$, and also taking into consideration \eqref{Phi_Psi_mu}, leads to
\begin{align}
  \frac{2}{r} \left(\Phi'+   \Psi' \right) =& 
 (2-\KB) \mu^2   e^{2\Psi}  \left(  e^{-\Phi}  \chi \qsol +e^{-2\Psi}A R'  -1\right)\left( \qsol  \frac{ e^{-\Phi}}{\chi} + \frac{  R' }{A} \right).
 \label{phipsi}
\end{align}
Notice in the above equation  $A$ appears in the denominator, thus, our solutions which follow cannot have $A=0$ over the whole of spacetime. However,
we do find such special solutions where $A=0$ and these are treated separately in section \ref{sec:solutions_special}.

In this work we consider scales that are much smaller than the MOND radius, and therefore, much smaller than $\mu^{-1}$. 
  This then implies that the RHS of \eqref{phipsi} is negligible due to the $\mu^2$ dependence, and we can safely assume
\begin{align}
\Psi = -\Phi,
\end{align}
since we can set the constant of integration to zero by a temporal coordinate transformation. 
We consider the two distinct cases separately: $\qsol=1$ and $\qsol=0$.

\subsection{Case $\qsol=1$} 
\label{sec:q1}
We first set $\qsol = 1$ into \eqref{reduced_scalar_SSS}  to find
\begin{align}
E = (1+\lambdas)\Qcal_0\left( e^{-\Phi}  A  +\chi R'\right),
\label{E_A}
\end{align}
and then add \eqref{reduced_Ein_11} and \eqref{reduced_Ein_AB} to obtain an equation for the potential $\Phi$ as the only variable, which is
\begin{align}
 \Phi'' + 2\Phi'  \left(\Phi' +  \frac{2}{r} \right)+   \frac{1}{r^2}   \left(1 - e^{-2\Phi}  \right)   =  0.
\label{leading_to_RN}
\end{align}
The unique solution to the above equation is the RN metric, that is,
\begin{align}
e^{2\Phi} = 1 - \frac{2\GN M}{r} + \frac{ \charge^2}{r^2},
\label{general_Phi_solution}
\end{align}
where we have defined the integration constants in analogy with the Einstein-Maxwell RN solution. Specifically, $M$ is the black hole mass and 
$\charge$ its charge.  The RN black hole has inner and outer horizons given by
\begin{align}
 \Rpm = \GN  M \left(1 \pm   \sqrt{1 -  \rcharge^2 } \right)
\end{align}
where $\rcharge \equiv \charge / (\GN M)$ is the normalized RN charge, and thus for a BH to exist we need $ -1 \le \rcharge \le 1$ with $|\rcharge| =1$ being the extremal case.


With this solution, taking the square root of both sides of \eqref{reduced_Ein_00} or \eqref{reduced_Ein_11} relates $E$ to the metric
\begin{align}
\frac{ \charge}{r^2} e^{-\Phi} =& \pm \sqrt{\nt} E,
\label{E_pm}
\end{align}
where it is understood that $\charge$ can take both positive and negative values. Notice that regardless of the sign of $\charge$, there remains an
ambiguity as to how the sign of $E$ is associated with the sign of $\charge$. Indeed, this is a matter of convention.  
To make a choice 
we use the electrodynamics analogy to define the vector field charge $\qA$ as
\begin{align}
  \qA  \equiv \lim _{r \rightarrow \infty} \frac{1}{4 \pi} \int_{S^2} F^{tr}  r^2 \sin \theta d \theta d \phi
=  - \lim_{r \rightarrow \infty}  e^{\Phi} E r^2 
=  \mp \frac{\charge}{\sqrt{\nt}},
\label{qA_def}
\end{align}
where the last equality is after using \eqref{E_pm}.  Thus, we choose the convention that $\qA$ and $\charge$ have the same sign, such that,
\begin{align}
    \qA \equiv   \frac{\charge}{\sqrt{\nt}}, 
\quad \mbox{and} \quad
  E =& -\frac{ \qA e^{-\Phi} }{r^2} 
.
\label{q_A_def}
\end{align}

At this point, we may determine the solution for the fields $A$ and $\phi$. 
Formally integrating \eqref{eq_def_E} we find $ \chi  =  e^{-\Phi} \left(\chi_0 + \frac{\qA}{ r} \right)$
where $\chi_0$ is an integration constant. 
The integration constant $\chi_0$ can be fixed by the frame freedom mentioned in section \ref{sec:asymptotics} and expanded on in Appendix-\ref{sec:appendix_lorentz}. Taking the $r\rightarrow \infty$ limit, we have that
$\chi\rightarrow \chi_0$, from which  $A_\mu^{(\infty)} =  (- \chi_0 , \sA\sqrt{\chi_0^2 - 1}  , 0,0)$, where $\sA = \pm 1$ is the sign of the 
spatial component of the vector field. Furthermore, from \eqref{q_A_def} we  have that $E\rightarrow 0$ and then  \eqref{E_A} implies that $R' \rightarrow -\sA\sqrt{1-1/\chi_0^2}$,
so that $\nabla_\mu \phi|^{(\infty)} = (  \Qcal_0 , -\sA \Qcal_0 \sqrt{\chi_0^2 - 1} /\chi_0, 0,0)$. This has the same form as
\eqref{coincide_frame_explicit} and thus, the two frames specified by $A_\mu$ and $\nabla_\mu \phi$ are coincident.
 Therefore, the integration constant $\chi_0$ has no physical meaning and we may set $\chi_0  = \pm 1$.
Given this last condition, we must also ensure that $\chi$ is never zero at any finite distance $r$, otherwise a pathology would occur. This is readily done 
by associating the constant $\chi_0$ with the sign of $\qA$, that is $\chi_0 = 1 $ corresponds to $\qA>0$ and $\chi_0 = -1 $  to $\qA<0$. Thus, in all generality we have that
\begin{align}
 \chi  =&  \chi_0 e^{-\Phi} \left(1 + \frac{|\qA|}{ r} \right),
 \label{chi_sol_gen}
\end{align}
Backgracing the steps until \eqref{E_pm}, the convention is that positive $\qA$ associates to a future directed $A^\mu$ and negative to past-directed $A^\mu$.

From \eqref{chi_sol_gen}  and  \eqref{chi_def} we solve for $A$ to find
\begin{align}
  A =&  
     \sA e^{-2\Phi} \sqrt{ \frac{2(|\qA| + \GN M)}{ r} + \frac{(1-\nt)\qA^2}{ r^2} }
\label{A_sol_spec}
,
\end{align}
where $s_A =\pm 1$ is the sign of $A$, resulting from taking the square root.
The vector field $A$ is well defined for all values of $r$ and has singularities exactly at the BH horizon at $\Rpm$ due to the $e^{-2\Phi}$ factor.
Just as in the case of the RN metric written in these coordinates, this is a coordinate and not a physical singularity, as we show more explicitly in
subsection \ref{sec:EF_regularization}.

Remember that $A_\mu \rightarrow - A_\mu$ is a symmetry of the action in the case that $\mu=0$, a condition which  we have already imposed, see section-\ref{sec:discrete_symmetry}.
In component form, this amounts to the symmetry under the simultaneous transformation  $\chi_0 \rightarrow - \chi_0$ and $s_A \rightarrow - s_A$,
in other words, $\{\chi_0 = 1 (\qA>0), s_A= -1\}$ and $\{\chi_0 = -1 (\qA<0), s_A= 1\}$  correspond to the same physical solution. Likewise,
 $\{\chi_0 = 1 (\qA>0), \sA= 1\}$ and $\{\chi_0 = -1 (\qA<0), s_A= -1\}$ is another set of equivalent solutions. Thus, we have two types of solutions and
without loss of generality, we fix $ \chi_0 = 1$, and let $s_A$ take both positive and negative values. To make things 
more precise, we set $\sA= - \epsilon_A \Sign{\qA}$, and let $\epsilon_A = \pm1$ be a parameter which specifies the type of solution.

Having found the form of $A$, we then determine $R'$ using \eqref{E_A} and \eqref{A_sol_spec} to get
~\footnote{Notice that $R=0$ 
is not a consistent solution because in that case we find that  $A = -\frac{\qA}{(1+\lambdas)\Qcal_0 r^2}$ 
which results in $\chi =  \chi_0 \sqrt{1 + e^{2\Phi} \frac{\qA^2}{(1+\lambdas)^2\Qcal_0^2 r^4}  } $. However,
 this $\chi$ solution cannot be consistent with \eqref{chi_sol_gen}.
}
\begin{align}
 R' =&  -\frac{1}{1 + \frac{|\qA|}{r}} \left[ \frac{1}{  (1+\lambdas)\Qcal_0 } \frac{ |\qA|  }{r^2}
-  \epsilon_A    e^{-2\Phi} \sqrt{ \frac{2(|\qA| + \GN M)}{ r} + \frac{(1-\nt)\qA^2}{ r^2} }  
\right]
\end{align}
One can further integrate the above equation to get $R(r)$, however, the result is rather complicated and not particularly illuminating and so we refrain of showing it except 
in the special case of $\charge=0$ below. We have thus found the full solution which depends on three parameters, being the mass $M\ge 0$, charge $\qA$ and discrete parameter
$\epsilon_A = \pm1$. 

We finally discuss the scalar $\Qcal$ which should neither diverge nor vanish within  the regime of validity of the solution, that is, for $r\ge \Rp$. We find
\begin{align}
\Qcal =&  \frac{\Qcal_0}{1 + \frac{|\qA|}{ r}  } \left[   1 
 +  \frac{\epsilon_A |\qA|}{ (1+\lambdas)\Qcal_0  r^2}    \sqrt{ \frac{2(|\qA| + \GN M)}{ r} + \frac{(1-\nt)\qA^2}{ r^2} }
\right].
\end{align}
We see that $\Qcal$ is never divergent for any $\epsilon_A=\pm1$, and moreover, in the case of the solution branch $\epsilon_A=1$ it never vanishes.
However, it can vanish at a non-zero $r$ in the case $\epsilon_A=-1$, and this can be avoided provided this point occurs at a point below the horizon. Indeed,
 starting from a set of parameters for the theory $\{\lambdas,\Qcal_0,\KB\}$ as well as (normalized) charge $\rcharge$, there is a lower bound on possible masses 
of black holes with this particular solution branch given by
\begin{align}
 \GN^2 M^2  > \frac{ \rcharge^2 
  \left[  2 \left( \frac{|\rcharge|}{\sqrt{\nt}} + 1\right) \left( 1 \pm   \sqrt{1 -  \rcharge^2 }  \right)
+   \rcharge^2 \left( \frac{1}{\nt} - 1\right)  \right]
 }{ \nt    (1+\lambdas)^2\Qcal_0^2   \left[ 1 \pm   \sqrt{1 -  \rcharge^2 }  \right]^6 }.
\label{eq:Qcal_epsilon}
\end{align}
Notice that this bound  does not include the $M=0$ case. This may be an indication that these solutions can only form with masses higher than this minimum mass bound. 
Ultimately, it is a question of BH formation rather than constraints on the theory parameters. The full viability of these solutions 
could be differentiated by a robust treatment of stability via perturbations, which we leave for further work.

\subsubsection{Schwarzschild case in suitable coordinates}
In the Schwarzschild case  $\charge=0$, the fields simplify significantly. Specifically we have that  $\chi = e^{-\Phi}$ and $E=0$, while
\begin{align}
  A =&  \sA e^{-2\Phi} \sqrt{ \frac{2\GN M}{ r}}, \qquad&  R' =& -A,
\end{align}
and the scalar field can be integrated to
\begin{align}
 \phi  =&  \Qcal_0 t  -2 \sA \Qcal_0 
\left[
 \sqrt{ 2 \GN M r}
-  \GN M \ln \left|\frac{ 1 + \sqrt{\frac{2 \GN M}{r}}  }{1 - \sqrt{\frac{2 \GN M}{r}}  }  \right|
\right].
\label{phi_Schwarzschild}
\end{align}
More interestingly $\Qcal=\Qcal_0$ exactly, which is a finite constant. This means that we retain the sign freedom in choosing $\sA$ in this case.
 This is because, in the Schwarzschild case  a new symmetry emerges due to the fact that $E=0$.
This allows transformations of the form $A\rightarrow -A$ and $R'\rightarrow-R'$ while keeping $\chi>0$  which leave the spherically symmetric field equations invariant,
 letting the vector field switch between ingoing and outgoing configurations irrespective of the original $A_\mu$ symmetry. 
This means that we can simply chose $s_A = 1$ by convention in the Schwarzschild case.

It is evident that both the vector field and the scalar field are divergent at the horizon. However, this is not a physical divergence. Observe that
\eqref{phi_Schwarzschild} allows us to define a new time coordinate $\tau(t,r)$ via $\phi = \Qcal_0 \tau$. More detailed,
we may transform to coordinates $\tau$ and $\rho$ via the coordinate transformation
$t(\tau,\rho) = \tau  +   2 \sA \left[\sqrt{  2 \GN M r} +  \GN M \ln\frac{ 1- \sqrt{\frac{2\GN M}{r}} }{1+    \sqrt{\frac{2\GN M}{r} } }  \right]$
and $r(\tau,\rho) =  \left(2\GN M\right)^{1/3}  \left(\frac{3}{2}|\tau - \rho|\right)^{2/3}$, which brings the metric in the
Lema\^{i}tre-Novikov form
\begin{align}
ds^2 = -d\tau^2 + \frac{2\GN M}{r} d\rho^2 + r^2 d\Omega
\end{align}
while the vector field aligns with the $\tau$ direction exactly, that is,  $ A_{\mu'} = (-1,0,0,0)$.~\footnote{
This form of the Schwarzschild solution in AeST theory has been independently discovered by T. Zlosnik as an educated guess that the BH should be continuously joined to FLRW 
solution for which $\phi=\Qcal_0 \tau$.} These coordinates acquire a physical meaning, as the rest-frame coordinates of a radially
ingoing ($\sA=1$) or radially outgoing ($\sA=-1$) geodesic observer freely falling towards to (or outwards from) the BH.

\subsubsection{Eddington-Finkelstein coordinates and regularization of RN case}
\label{sec:EF_regularization}
The singularity of the RN metric occuring at $r = \Rpm$ is commonly removed by transforming to
Eddington-Finkelstein (EF) coordinates. We show in this section that transforming to EF coordinates also removes the singularities of the vector and scalar field.

As in GR, we define the tortoise radial coordinate $\rEF$ using
$\frac{d\rEF}{dr} \equiv  e^{-2\Phi} =  \frac{1}{\left(1-\frac{\Rp }{r}\right)\left(1-\frac{\Rm}{r}\right)}$
leading to
\begin{align}
 \rEF & = r + \frac{ \Rp^2 }{\Rp- \Rm} \ln \left|\frac{r}{\Rp}-1\right|+\frac{\Rm^2}{ \Rm - \Rp} \ln \left|\frac{r}{\Rm}-1\right|
\end{align}
which is then used to define the ingoing $u_-$ and outgoing $u_+$ EF coordinates via $u_{\pm} =  t \pm \rEF$. In these coordinates the metric takes the well known form
$  ds^2 = - e^{2\Phi} du_+ du_-  + r^2 d\Omega^2$.
But the above metric, while non-singular at the horizon, is actually degenerate: it's determinant vanishes at the horizon, and in fact its inverse becomes singular there.  
In GR this is usually resolved  be further transforming to Kruskal coordinates, however, another possibility is to keep within either the ingoing or outgoing branch of the EF
coordinates. In that case, we transform $t$ to either $u_+$ or $u_-$ but keep $r$ as a coordinate so that $ds^2 = - e^{2\Phi} du_{\pm}^2 \pm 2  du_{\pm}  dr + r^2 d\Omega^2$
which is non-singular at the horizon and not degenerate. We now demonstrate that the vector and scalar fields are also non-singular at the horizon
when expressed in this coordinate system.


 Using $2\GN M = \Rp + \Rm  $ and $q_A^2\nt  = \Rp\Rm $ we can express the vector field as
\begin{align}
    \covec{A} & =
 -\left( 1 +   \frac{ \sqrt{\Rp\Rm}}{\sqrt{\nt} \, r} \right)   du_{\pm}
 +  \frac{ \sEF \left( 1 +  \frac{ \sqrt{\Rp\Rm}  }{\sqrt{\nt}\,r} \right)
+ \sA \sqrt{ 
\frac{\Rp+\Rm}{r} 
+    \frac{2\sqrt{\Rp\Rm}}{\sqrt{\nt} \, r} 
+  \frac{1-\nt}{\nt} \frac{\Rp \Rm}{r^2} 
} }{ \left( 1 - \frac{\Rp}{r} \right)\left( 1 - \frac{\Rm}{r} \right) }
dr,
\end{align}
where $\sEF = 1$ if we choose the outgoing coordinate $u_+$ and $\sEF = -1$ for the ingoing $u_-$ case. 
 We proceed by expanding the radius $r$ near the horizon(s). We first set $r = \Rp (1 + \epsilon)$ and consider the limit $\epsilon \rightarrow 0$.
Provided that $\sA = -\sEF$, we find
\begin{align}
    \covec{A} & =
 -\left( 1 +   \sqrt{\frac{\Rm}{\nt\Rp} } \right)   du_{\pm}
+ \sEF \frac{dr}{2 \left(1 +  \sqrt{\frac{\Rm}{\nt\Rp}} \right)  } 
+ O(\epsilon),
\label{eq_EF_A}
\end{align}
which is finite, while a divergence occurs as $\epsilon\rightarrow 0$ if $\sA = \sEF$. We analogously investigate the inner horizon and set  $r = \Rm (1 + \epsilon)$,
leading to a similar non-divergent result as in \eqref{eq_EF_A}, except that $\Rm\leftrightarrow\Rp$ in how they occur in \eqref{eq_EF_A}, 
and for the same correspondence between $\sEF$ and $\sA$.
We conclude that the field divergences in $A_\mu$ and $\phi$ at the RN horizons  
are a coordinate effect, just as in the case of the metric, and can be removed with a suitable coordinate transformation.

\subsection{Case $\qsol=0$}
\label{sec:q0}
In this case, \eqref{scalar_charge} does not necessarily imply that $\phi_0=0$, which means that 
$\phi_0$ can in principle take any value. Specifically, in this case we have that 
 \begin{align}
   E = (1+\lambdas)\Qcal_0 \chi R' + \frac{\phit_0  e^{-2\Phi} }{\chi r^2},
\label{E_A_q0}
\end{align}
where $\phit_0 \equiv \phi_0 / \Qcal_0$, and which is to be contrasted with \eqref{E_A} from the $\qsol=1$ case.
  Setting $\qsol=0=\mu$ and $\Psi=-\Phi$ into \eqref{reduced_Ein_00} and \eqref{reduced_Ein_AB}, and eliminating $R'$ in terms of $E$ and $\phi_0$ using \eqref{E_A_q0},
leads to 
\begin{align}
   \frac{2}{r}   \Phi'  + \frac{1-e^{-2\Phi}}{r^2} =& - \nt   E^2 + \frac{ 2 - \KB}{2(1+\lambdas)}\frac{\phit_0^2  e^{-4\Phi} }{\chi^2 r^4},
 \label{reduced_Ein_00_q0}
\\
 \Phi'' + 2 \Phi'  \left(\Phi' +  \frac{1}{r} \right) =&  \nt E^2 - \frac{(2-\KB) }{2(1+\lambdas) }     \frac{\phit_0^2  e^{-4\Phi} }{\chi^2 r^4} ,
\end{align}
for  \eqref{reduced_Ein_00} and \eqref{reduced_Ein_AB} respectively. 
Therefore, adding the last two equations, leads once more to  \eqref{leading_to_RN} as in the $\qsol=1$ case, and so once again 
the RN metric \eqref{general_Phi_solution} is the  most general solution.

Given the metric solution  \eqref{general_Phi_solution}, we insert it back into \eqref{reduced_Ein_00_q0} to find the
relation 
\begin{align}
 E =& \sE \frac{e^{-\Phi}}{r^2} \sqrt{  \frac{ \charge^2 }{\nt} +  \frac{\mt^2 \phit_0^2  e^{-2\Phi} }{\chi^2} },
\label{E_chi_Phi_q0} 
\end{align}
where $\sE = \pm 1$. Then, directly integrating  \eqref{eq_def_E} and eliminating $E$ using this last relation, leads to
 $e^{2\Phi} \chi^2 = \frac{\nt \mt^2}{\Qcal_0^2\charge^2} \left[\left( u_0 -\sE \frac{\Qcal_0 \charge^2 }{\mt\nt} \frac{1}{r} \right)^2  -  \phi_0^2 \right]$,
where $u_0$ is an integration constant. Once more we apply the frame freedom which fixes $u_0^2  =  \varphi_0^2  + \frac{ \Qcal_0^2\charge^2 }{\nt \mt^2 }$,
and so the solution $\chi$ turns into
\begin{align}
 \chi =&  \schi e^{-\Phi} \sqrt{1 + \frac{2 \su \mt |u_0| }{ \Qcal_0 }  \frac{1}{r}  + \frac{  \charge^2  }{ \nt}  \frac{1}{r^2}   },
\label{chi_sol_q_1_general}
\end{align}
where $\schi=\pm1$ and we have set $u_0 = - \su \sE|u_0|$, where $\su = \pm 1$ as the two possibilities relating to the sign (but not equal) of $u_0$.
The first case, $\su = +1$, implies that $\chi$ is well defined for all values of $r$ and diverges only at the two horizons and at the 
singularity $r=0$. The second case, $\su = -1$, can lead to $\chi$ vanishing at a finite distance $r>\Rpm$. A necessary and sufficient condition for
avoiding this is 
\begin{align}
    \GN M   \ge \frac{\mt |u_0|}{\Qcal_0},
\label{u_0_A_condition}
\end{align}
which is to be applied only to the $\su = -1$ case and not to $\su = 1$. This last relation is to be contrasted with \eqref{eq:Qcal_epsilon} in the case of the $\qsol=1$ solution,
only in the latter case, the bound comes from requiring the non-vanishing of $\Qcal$.

We now give a physical meaning to the constant $u_0$ appearing above.  Using \eqref{chi_sol_q_1_general}  into \eqref{E_chi_Phi_q0} leads to
\begin{align}
  E=&\sE  \schi \frac{e^{-\Phi}}{r^2} \frac{ - \sE   \frac{\mt u_0}{\Qcal_0} +  \frac{\charge^2 }{\nt r} 
  }{ \sqrt{1 + \frac{2 \su \mt |u_0| }{ \Qcal_0 }  \frac{1}{r}  + \frac{  \charge^2  }{ \nt}  \frac{1}{r^2}    } }, 
  \label{E_q_1_sol}
\end{align}
and analogously to the $\qsol=1$ case, we  define an $A$-charge $\qA$ as
\begin{align}
  \qA  \equiv \lim _{r \rightarrow \infty} \frac{1}{4 \pi} \int_{S^2} F^{tr}  r^2 \sin \theta d \theta d \phi
=  - \lim_{r \rightarrow \infty}  e^{\Phi} E r^2  =   \schi \frac{\mt u_0}{\Qcal_0} .
\end{align}
Notice that just as in the $\qsol=1$ case, there is an ambiguity as to how we associate charge to $u_0$. Taking the limit $\phi_0\rightarrow 0$ we find
$\qA \rightarrow -  \schi  \sE \frac{\charge }{\sqrt{\nt}}$, and so, we choose the convention that in this limit
$\qA \rightarrow  \frac{\charge }{\sqrt{\nt}}$ as in \eqref{q_A_def} from the $\qsol=1$ case. This fixes $\schi =1$,  while $\sE=-1$ so that $u_0 = \su |u_0|$. Then
\begin{align}
  \qA  =&  \frac{\mt u_0}{\Qcal_0},
   \label{q_A_def_q_1}
\end{align}
which implies the following relation between the $A$-charge, scalar charge $\phi_0$ and BH charge $\charge$,
\begin{align}
  \qA^2  =&  \frac{\mt^2}{\Qcal_0^2}  \varphi_0^2  + \frac{\charge^2 }{\nt  }.
   \label{q_0_charge_relation}
\end{align}

With the above considerations and conventions we form the final solution for $\chi$ and $E$ as
\begin{align}
\chi =&   e^{-\Phi} \sqrt{1 + \frac{2\qA}{r}  + \frac{  \charge^2  }{ \nt}  \frac{1}{r^2}   },
 \quad  & E =& - \frac{e^{-\Phi}}{r^2} \frac{   \qA +  \frac{\charge^2 }{\nt r} }{ \sqrt{1 + \frac{2 \qA}{r}  + \frac{  \charge^2  }{ \nt}  \frac{1}{r^2}    } },
\label{chi_q_0_complete}
\end{align}
while from $\chi$, using \eqref{chi_def} we immediately obtain $A$ as
\begin{align}
   A =&  \sA e^{-2\Phi} \sqrt{  \frac{2 (\GN M +\qA)}{r}  + \frac{  \charge^2 (1 - \nt)  }{ \nt}  \frac{1}{r^2}    }.
   \label{eq_A_q_0_sol}
\end{align}
This is in fact identical to \eqref{A_sol_spec}, even though the solution for $\chi$ is different (as well as the relation between $\qA$ and $\charge$). 
Likewise, using \eqref{chi_q_0_complete}  into \eqref{E_A_q0}  and integrating, give the solution for the scalar as
\begin{align}
(1+\lambdas)\phi =& \ln \sqrt{  1  +\frac{2 \qA }{r}  +\frac{ \charge^2  }{ \nt }  \frac{1}{r^2}}
+  \frac{ s_{\phi_0} }{2 \mt  } 
  \ln\frac{ 1 + \frac{\qA + \mt |\phit_0| }{ r}  }{ 1 + \frac{\qA - \mt |\phit_0|   }{ r }   },
\label{eq_phi_q_0_sol}
\end{align}
where $s_{\phi_0}$ is the sign of $\phi_0$.

Before concluding this section, note that important scalar $\Qcal = \Qcal_0 e^{2\Phi}AR'$ behaves correctly for all branches. 
Moreover, we note that all the above expressions are valid also in the cases of $\charge=0$, or if $\phi_0=0$ as we have explicitly verified,
and so these special cases connect smoothly to the general case. In appendix \ref{app:q_0_Schwarzschild} we show the special Schwarzschild case where $\charge=0$.

\section{Algebraically special solutions with $A=0$}  
\label{sec:solutions_special}
In section \ref{sec:solutions_general} we have assumed that $A\ne0$ which is important in reaching \eqref{phipsi}.
Let us now consider  algebrically special cases, which amount to setting $A=0$ everywhere in \eqref{eq_def_E},
 \eqref{reduced_scalar_SSS}, \eqref{scalar_charge}, \eqref{reduced_Ein_00}, \eqref{reduced_Ein_11} and \eqref{reduced_Ein_AB}. 
This is a case partially motivated by similar type of solutions found in the case of TeVeS theory~\cite{giannios2005spherically,sagi2008black}. 
Note that this implies that $\chi = \pm 1$ and so \eqref{eq_def_E} becomes $E= \chi \Phi' = \pm \Phi'$. 
We consider both $\qsol=1$ and  $\qsol=0$ separately.

\subsection{Case $\qsol=1$} 
\label{A0section}
Here \eqref{scalar_charge} implies that $\phi_0=0$, thus, setting $A=0$ and $\qsol=1$ in  
\eqref{reduced_scalar_SSS},  \eqref{reduced_Ein_00}, \eqref{reduced_Ein_11} and \eqref{reduced_Ein_AB}, 
and leaving the details for appendix-\ref{sec:appendix_A_special_q} we find the scalar field as 
\begin{align}
\phi  =&  \Qcal_0 t + \frac{1}{2n (1+\lambdas)} \ln \frac{|u-u_2|}{|u-u_1|} 
\label{phi_special_A_1}
\end{align}
and the metric as 
\begin{align}
 ds^2 = -  \frac{|u-u_2|^{\frac{1}{n}}}{|u-u_1|^{\frac{1}{n}}}  dt^2  + \frac{r_0^2}{u^2}  \frac{|u-u_1|^{\frac{1+n}{n}}}{|u-u_2|^{\frac{1-n}{n}} } 
  \left[ 
  \frac{ du^2}{ \nt (u - u_1) (u - u_2)  u^2}  +   d\Omega^2 \right].
\label{g_special_A_1}
\end{align}
The new coordinate $u$ is implicitly given through the coordinate transformation 
\begin{align}
r(u)  &=    \frac{r_0}{|u|} 
  \frac{|u-u_1|^{\frac{1+n}{2n}}}{|u-u_2|^{\frac{1-n}{2n}} },
\label{r_u_solution}
\end{align}
where $r_0$ is an arbitrary scale, $n \equiv \sqrt{1 - \nt}$
and $u_1 = -1/(1+n)$ and $u_2= -1/(1-n)$ are the two roots of the polynomial $p(x)=1+2x+ \nt x^2$,  and since $0\leq n < 1$ then $u_2<u_1<0$. The scalar curvature of this solution is
\begin{align}
 \Rcal
 =& -   \frac{2u^4}{r_0^2(u - u_1) (u - u_2) }     \frac{|u-u_2|^{\frac{1-n}{n}} }{|u-u_1|^{\frac{1+n}{n}}} 
\end{align}

From \eqref{r_u_solution} we see that there are four disconnected solution branches:
\begin{itemize}
\item Branch I: $u\ge 0 $. This branch covers the manifold from a minimum radius $r=r_0$ corresponding to $u\rightarrow \infty$, to spatial infinity $r\rightarrow \infty$ corresponding to $u=0$. The scalar curvature is finite and negative during the entire range, corresponding to $\Rcal=0$ as $r\rightarrow\infty$ and $\Rcal \rightarrow - 2/r_0^2$ as $r\rightarrow r_0$. 
Moreover, as $r\rightarrow\infty$ we have that $\phi' \rightarrow 0$, and so the frame of $A_\mu$ and $\nabla_\mu\phi$ coincide.
This interesting behaviour deserves further investigation elsewhere. 

We may also define an $A$-charge for this solution analogously to section \ref{sec:solutions_general} by setting 
$ \qA  \equiv \lim _{r \rightarrow \infty} \frac{1}{4 \pi} \int_{S^2} F^{tr}  r^2 \sin \theta d \theta d \phi
=  \mp \lim_{u \rightarrow 0^+}  e^{\Phi} u r$. This leads to $\qA= \mp r_0 / \sqrt{\nt}$. 
\item Branch II: $u_1<u\le 0$. For this branch, the upper limit corresponds to spatial infinity $r\rightarrow\infty$, while the lower limit 
corresponds to $r=0$, however, there the scalar curvature diverges. Thus this branch has a naked singularity and is not
expected to be physical.
\item Branch III : $u_2 < u < u_1$. This branch has a metric signature  $--++$ and is therefore not physical.
\item Branch IV: $u<u_2$.  This last branch has some similarities with branch I, that is, as $u\rightarrow -\infty$, then $r\rightarrow r_0$ while 
as $u\rightarrow u_2$ then $r\rightarrow \infty$.  
The former limit also has the property that the scalar curvature $\Rcal \rightarrow -2/r_0^2$, however,
 the latter limit leads to three subcases concerning the behaviour of $\Rcal$. If $0\le n < 1/2$ (branch ${\rm IV}_{\rm a}$) then, 
$\Rcal\rightarrow 0$ as $r\rightarrow \infty$, and this behaviour is similar to branch I. If $1/2 < n < 1$ (branch ${\rm IV}_{\rm c}$) 
then  $\Rcal$ diverges as $r\rightarrow \infty$, and so this subcase is pathological. If $n=1/2$ exactly (branch ${\rm IV}_{\rm b}$), then, $\Rcal$ 
interpolates between its value at $r=r_0$ and $\Rcal = - \frac{81}{8r_0^2}$ at $r\rightarrow \infty$.  
Nevertheless, as $u\rightarrow u_2$ ( $r\rightarrow \infty$), we have that $\nabla_\mu \phi \nabla^\mu\phi$ diverges. 
\end{itemize}

Upon inspection, Branch I and Branch IV seems to cover two portions of the Eling-Jacobson wormhole~\cite{Eling:2006df} 
which was first discovered in Einstein-Aether theory, where the solution is joined between $u=\infty$ to $u=-\infty$ through $r_0$.~\footnote{See appendix-\ref{sec:appendix_A_special_q} and 
specifically the footnote below \eqref{Psi_equation_A_0} and \eqref{Phi_equation_A_0}. The existence of Eling-Jacobson wormhole~\cite{Eling:2006df,Oost:2021tqi} 
  solutions in AeST theory  has been brought to our attention by William Barker, who has discovered them in another work with collaborators~\cite{Wormhole_TBS}.}.


\subsection{Case $\qsol=0$}
\label{sec:A0q0}
We briefly comment on the subcase $\qsol=0$, and again leaving the details for Appendix-\ref{sec:appendix_A_special_q}, we get a condition $\mu R'=0$ 
from combining the \eqref{reduced_scalar_SSS}, \eqref{reduced_vector_SSS} and \eqref{G_01_SSS} equations. 
It turns out that the $R'=0$ possibility reduces to the $A=0, q=1$ case considered above,  
with the identification $\nt \rightarrow \KB/2$.  However, here we also have a non-zero $\phi_0$ and this leads to the relation
 $\Qcal_0 r_0  =   |\phi_0|  \sqrt{\frac{\KB}{2}}$ while $\Sign{u} = \Sign{\phi_0}$, that is, the branch is chosen by the sign of $\phi_0$. 

Letting $\mu=0$ without a priori setting $R'$ to anything leads to a single equation for the metric potential 
\begin{align}
  \Psi'' 
+ \frac{1}{r} \left( 3e^{2\Psi}  + 1 \right)  \Psi' 
+ \frac{1}{r^2} e^{2\Psi} \left(  e^{2\Psi}  - 1 \right) 
  =& 0,
\label{eq_A_0_q_0_mu_0_Phipp}
\end{align}
which admits a Schwarzschild solution for the metric, and  scalar hair
\begin{align}
\phi =  \frac{1 \pm  \sqrt{\frac{2 + \KB \lambdas}{2-\KB}} }{2  (1+\lambdas)  } \ln \left(1 - \frac{2 \GN M}{r} \right).
\end{align}
coming from solving \eqref{Rp_q_0_mu_0_app}. 
Interestingly, satisfying the Einstein equation \eqref{G_00_q_0_mu_0_app}, leads to the constraint
\begin{align}
    \phi_0   =& \pm\sqrt{\frac{2 + \KB \lambdas}{2-\KB}}  \GN M.
\end{align}
which removes any explicit dependence on $\phi_0$ from this solution.
Unfortunately we have not found more general solutions, and given the low interest in this particular case, we leave it for another investigation.

\section{Discussion}
\label{sec:discussion}
We have already validated that the scalar $\Qcal$  is regular at the BH horizons in section-\ref{sec:solutions_general}. Moreover,
since the metric solution is of the RN type, then all curvature invariants are also well-behaved. In the case $\qsol=1$ we have
 that $\Ycal =  -e^{\Phi} E / (1+\lambdas)  = -\frac{1 }{ 1+\lambdas } \frac{ \qA  }{r^2}$ which is therefore regular at the horizon 
and the same can be shown to true in the $\qsol=0$ case, where $ \Ycal = \Qcal_0^2   e^{2\Phi}\chi^2 {R'}^2$, using \eqref{chi_q_0_complete} and \eqref{E_A_q0}.
Hence, the scalar kinetic term is also regular there.

Another important scalar that frequently appears in shift-symmetric theories is the  term  $S^\mu S_\mu$ formed out of the shift symmetry Noether current \eqref{scalar_eq}.
If $\Phi=-\Psi$, then a commonly used argument  (see \cite{creminelli2020hairy} as a coherent example) 
states that the horizon should be a regular locus and any scalar quantity $\Ocal$  should not diverge there. Otherwise, 
given   $\Ocal$ which diverges at the horizon, adding a term $\epsilon_{h} \Ocal$ for a constant $\epsilon_{h}$ to the Lagrangian 
would change the solution to the field equations for any $\epsilon_{h}$, regardless of how small $\epsilon_{h}$  is made.
 For the $\qsol=1$ branch, this term is identically zero, while for the $\qsol=0$ branch,
it evaluates to $S^\mu S_\mu = \frac{\phit_0^2(2-\KB)^2}{e^{2\Phi}\chi^2 r^4}$, which for the  \eqref{chi_q_0_complete} solution it is regular 
 at all points $r >0$, including the  horizons $r=\Rpm$.

Checking the divergence of scalar quantities at horizons becomes even more involved if we require that the BH should be continuously joined to a cosmological spacetime.
The $q=1$ case of section \ref{sec:solutions_general} has the right properties for doing so, however, as we have explicitly removed the cosmological dependence (the solutions are 
asymptotically flat), more checking is necessary for ensuring this is the case. One could try to find solutions which are asymptotically de Sitter, see 
\cite{Babichev:2024txe} where this was addressed in the case of scalar Gauss-Bonnet theory, and check that they do remain regular at both the BH and at the de Sitter horizon.
We conjecture that this does not pose a problem, and leave it for a  future investigation.

Finally, let us consider what could happen if the full $\Jcal(\Ycal) = \Fcal(\Ycal,\Qcal_0)/(2-\KB)$ function is included, so that the solution smoothly joins to the MOND regime.
Upon inspection of the Einstein equations, it seems that in the limit $\mu\rightarrow 0$, that is, considering scales smaller 
than $r_C \sim (\GN M/a_0)^{1/6} \mu^{-2/3}$ (see~\cite{Verwayen:2023sds}), then the condition $\Psi = -\Phi$ is retained. However, the RHS of \eqref{leading_to_RN} is no longer 
expected to be zero, but rather proportional to 
$\Ycal   \Jcal_{\Ycal} - \Jcal$, hence, the full solution will no longer be the RN solution, once scales close to the MOND radius become important.
Interestingly, since $E = 0$ in the ($\qsol=1)$ Schwarzschild case, then also $\Ycal = 0$, and hence, we do not expect the Schwarzschild solution of section \ref{sec:q1} 
to be smoothly joinable to the MOND regime.
These issues could be addressed elsewhere.

\section{ Conclusion}
\label{sec:conclusion}
We have found the most general classes of static spherically symmetric solutions in the AeST theory 
assuming the strong field regime which amounts to looking at scales smaller than the characteristic radius where MOND effects would appear.
We found two classes of stealth black holes -- black holes with geometries identical to GR black holes -- with non-trivial secondary hair. 
Specifically, the BH geometry is that of the Reissner-Nordstrom spacetime.
The first class, $\qsol=1$, was presented in section-\ref{sec:q1} and consists of consistent black hole candidates with zero shift-charge, 
which can in principle be smoothly connected to the cosmological regime. This is possible because the gradient of the scalar field for this class is by construction timelike.
Some further important considerations before the smooth extension of this class to cosmology can be established are found in the discussion section-\ref{sec:discussion}.
The second class,  $\qsol=0$,  was presented in section-\ref{sec:q0} and consists of consistent black hole candidates with non-zero shift-charge and 
spacelike scalar field  gradient. As such, it is unlikely that this class can be smoothly connected  to cosmology.
Apart from the RN black holes, we also found  algebraically special solutions where the vector field is aligned with the time direction, see section-\ref{sec:solutions_special}. 
These solutions have no horizon and can be extended only down to a finite distance $r_0$. The solution space of these algebraically special  class also contains
non-physical branches which contain naked singularities. Our solutions are summarized in table-\ref{Tab_solutions}.

Since we have found the most general static spherically symmetric vacuum solutions, these can also be considered as possible exterior stellar solutions,
for example the neutron star solutions found in \cite{reyes2024neutronstarsaetherscalartensor}. Importantly, it would be interesting to study the stability of these solutions
to small time-dependent fluctuations. Ultimately, only the stable branches should be considered as physically viable solutions. Studying 
such fluctuations and resulting quasinormal modes and gravitational radiation, as well as resulting observational constraints on parameters of the theory is left for future work.
Finally, it would be very interesting to investigate the thermodynamics of our new solutions, another subject to be studied in further work.

\begin{acknowledgments}
We thank L. Blanchet, W. Barker, C. Charmousis, R. Emparan, G. Esposito-Far\`ese, E. Kiritsis, T. Mistele, I. Sawicki, T. Sotiriou,  L. Trombetta and T.G.Zlosnik for interesting discussions. 
This work was co-financed by the European Structural and Investment Funds and the Czech Ministry of Education, Youth and Sports (MSMT) of the Czech Republic 
 (Project FORTE – $\mathrm{PCZ.02.01.01/00/22\_008/0004632}$). D.V. thanks Pedro Ferreira, the BIPAC and the Oxford subdepartment of Astrophysics, 
for their warm welcome and hospitality during a research stays in spring in and autumn 2024.
 C.S. further acknowledges support by the Royal Society Wolfson Visiting Fellowship ``Testing the properties of dark matter with new statistical tools and cosmological data''.
\end{acknowledgments}

\begin{landscape}
\begin{table}
\begin{tabular}{|c|cc|c|c|}
\hline
Class &\multicolumn{1}{c|}{Subclass}& Solution & Parameters & Section \\ 
\hline \hline 
\multirow{14}{*}{$A\ne 0$}& \multicolumn{2}{c|}{
 Metric: $ds^2 = - e^{2\Phi} dt^2  +  e^{-2\Phi}  dr^2 + r^2 d\Omega^2$, $\qquad e^{2\Phi} = 1 - \frac{2\GN M}{r} + \frac{\charge^2}{r^2}$}& & \\  \cline{2-3} 
 
& \multicolumn{1}{l|}{\parbox{4cm}{
\begin{align}
q&=1\nonumber\\
\phi &= \Qcal_0 t + \Qcal_0 R(r)\nonumber
\end{align}}}  
&
\parbox{13cm}{%
\begin{align}
A &= 
\sA e^{-2\Phi} \sqrt{\frac{2(|\qA| + \GN M)}{r} + \frac{(1-\nt)\qA^2}{r^2}} \nonumber\\
R' &= - \frac{1}{1 + \frac{|\qA|}{r}} \left[ 
    \frac{1}{(1+\lambdas)\Qcal_0} \frac{\qA}{r^2} 
    + \epsilon_A e^{-2\Phi} \sqrt{\frac{2(|\qA| + \GN M)}{r} + \frac{(1-\nt)\qA^2}{r^2}}
\right]\nonumber
\end{align}
} 
 &  $M, \qA, \epsilon_A$ &\ref{sec:q1} 
\\ \cline{2-4}
 & \multicolumn{1}{c|}{$q=0$} & \parbox{13cm}{%
\begin{align}
   A =&  \sA e^{-2\Phi} \sqrt{  \frac{2 (\GN M +\qA)}{r}  + \frac{  \charge^2 (1 - \nt)  }{ \nt}  \frac{1}{r^2}    } \nonumber\\
(1+\lambdas)\phi =& \ln \sqrt{  1  +\frac{2 \qA }{r}  +\frac{ \charge^2  }{ \nt }  \frac{1}{r^2}}
+  \frac{ s_{\phi_0} }{2 \mt  } 
  \ln\frac{ 1 + \frac{\qA + \mt |\phit_0| }{ r}  }{ 1 + \frac{\qA - \mt |\phit_0|   }{ r }   }\nonumber
\end{align}
}
&   $M, \charge, \phi_0, \sA$ &\ref{sec:q0}   
\\ 
\hline
\hline
\multirow{7.4}{*}{$A=0$} &  \multicolumn{1}{c|}{$q=1$} & \parbox{13cm}{%
\begin{align}
ds^2 &= -  \frac{|u-u_2|^{\frac{1}{n}}}{|u-u_1|^{\frac{1}{n}}}  dt^2  + \frac{r_0^2}{u^2}  \frac{|u-u_1|^{\frac{1+n}{n}}}{|u-u_2|^{\frac{1-n}{n}} } 
  \left[ 
  \frac{ du^2}{ \nt (u - u_1) (u - u_2)  u^2}  +   \Sigma_{AB} dx^A dx^B\right] \nonumber\\
\phi  &=  \Qcal_0 t + \frac{1}{2n (1+\lambdas)} \ln \frac{|u-u_2|}{|u-u_1|} \nonumber
\end{align}
} 
& $r_0$ &\ref{A0section} 
\\ 
\cline{2-4}
  & \multicolumn{1}{c|}{$q=0$}& \parbox{13cm}{%
\begin{align}
e^{2\Phi} &= 1 - \frac{2\GN M}{r} \nonumber\\
\phi &=  \frac{1 \pm  \sqrt{\frac{2 + \KB \lambdas}{2-\KB}} }{2  (1+\lambdas)  } \ln \left(1 - \frac{2 \GN M}{r} \right)\nonumber
\end{align}
}
& $M$ &\ref{sec:A0q0}
 \\ 
\hline
\end{tabular}
\caption{Summary of all the types of solutions including field forms and parameters additional to the ones of the theory}
\label{Tab_solutions}
\end{table}
\end{landscape}

\appendix
\section{ Lorentz transformations and flat spacetime symmetries}
\label{sec:appendix_lorentz}
A general Lorentz transformation $\Lor{\mu}{\mu'}$ preserves the Minkowski metric, that is, if we assume a cartesian system such that
$\eta_{\mu\nu} = diag(-1,1,1,1)$ then $\eta_{\mu'\nu'} = \Lor{\mu}{\mu'}\Lor{\nu}{\nu'} \eta_{\mu\nu}  $ is also of the form $\eta_{\mu'\nu'} = diag(-1,1,1,1)$. 
The vector field $A_\mu$ and the scalar field gradient $\nabla_\mu\phi$ will also transform. 
We can boost into the frame of either one to simplify expressions, however, in general, these two frames do not necesserily have to coincide.

Consider general Lorentz boosts in a direction $\vec{\beta}$ 
\begin{align}
t' =& \gamma \left( t - \vec{\beta}\cdot \vec{x}\right),
\\
\vec{x}' =& \vec{x} + \frac{\gamma-1}{\beta^2} \left(\vec{\beta}\cdot \vec{x}\right) \vec{\beta} - \gamma t \vec{\beta},
\end{align}
where
\begin{align}
\gamma =& \frac{1}{\sqrt{1 - \beta^2 }},
\end{align}
and where $\beta^2 \equiv |\vec{\beta}|^2$.
Then, the vector and scalar fields transform as
\begin{align}
A_{0'} =& \gamma \left(A_{0} +  \beta^i A_i \right), \qquad & A_{i'} =&  A_i + \frac{\gamma-1}{\beta^2} \vec{\beta} \cdot \vec{A}  \beta_i + \gamma A_0 \beta_i ,
\\
 \frac{\partial\phi}{\partial t'} =& \gamma \left(  \frac{\partial\phi}{\partial t}+  \beta^i \partial_i \phi \right),
\quad
&
\grad_{i'} \phi =&  \grad_i \phi + \frac{\gamma-1}{\beta^2} \vec{\beta} \cdot \grad\phi   \beta_i + \gamma  \frac{\partial\phi}{\partial t}  \beta_i .
\end{align}

Now suppose that in a general frame our fields have the form
\begin{align}
A_{0} =& - \chi_0, \qquad & A_i =& \sA \sqrt{\chi_0^2 - 1}  \hat{z},
\\
\dot{\phi} =& \Qcal_0,  \quad & \grad_i \phi =&  \Qcal_0  \phi_f \hat{z},
\end{align}
where $\chi_0$ and $\phi_f$ are arbitrary constants, with the latter obeying $\phi_f <1$. We then perform a Lorentz transformation into a special frame chosen by
$\beta = \sA \sqrt{1 - 1 /\chi_0^2}$ which is none other than the vector frame, that is, $A_\mu' = (-1,0,0,0)$. With this choice of $\beta$, the scalar transforms to 
\begin{align}
\frac{\partial\phi}{\partial t'} =&  \chi_0 \Qcal_0  \left(1  +  \phi_f \beta \right),
\quad
& \grad_{i'} \phi =&  \chi_0  \Qcal_0 \left(  \phi_f     +     \beta \right) \hat{z}.
\label{coincide_frame}
\end{align}

Redefining $\Qcal_0$ and $\phi_f$ to $\Qcalh_0$ and $\phih_0$ we get the vector frame (performing also rotations to bring the $\hat{z}$ axis into a general direction), as
\begin{align}
A_0 =& -1 \qquad &  \vec{A} =& 0,
\\
\dot{\phi} =&  \frac{ \Qcalh_0}{  \sqrt{1  -  \phih_0^2 } } \qquad& \grad \phi =&   \frac{\Qcalh_0  \phih_0}{ \sqrt{1  -  \phih_0^2 } } \nhat,
\\
\phi =& \frac{ \Qcalh_0}{  \sqrt{1  -  \phih_0^2 } }  \left[t + \phih_0 r \right],
\end{align}
with $\phih_0<1$, s.t. $\nabla_\mu \phi \nabla^\mu\phi =  - \Qcalh_0^2<0$,
and where $r = \sqrt{x^2+y^2+z^2}$, is the distance along $\nhat$.
Likewise, in the scalar frame, we have
\begin{align}
A_{0} =& -\gamma \qquad&  \vec{A} =&   \gamma \phih_0  \nhat
\\
\dot{\phi} =&    \Qcal_0 \qquad& \grad \phi =&  0
\end{align}
Thus, in general an additional parameter appears, $\phih_0$, which expresses the tilt of the two frames: scalar vs vector.

Suppose, however, that we have a situation where $\phi_f + \beta =0$, in \eqref{coincide_frame}, so that
\begin{align}
\dot{\phi} =& \Qcal_0  \quad & \grad_i \phi =&  - \Qcal_0  \sA \frac{\sqrt{\chi_0^2 - 1}}{\chi_0} \hat{z}.
\label{coincide_frame_explicit}
\end{align}
Then the two frames coincide, such that
\begin{align}
\frac{\partial\phi}{\partial t'} =&  \Qcal_0/\chi_0 \quad & \grad_{i'} \phi =& 0
\label{coincide_frame_transform}
\end{align}
This is indeed the case of our solutions found in section-\ref{sec:solutions_general}. 

\section{Important middle calculation steps}
\label{sec:appendix_calc}
Using the metric \eqref{eq_SSS_metric}, where to remind the reader, capital Latin letters denote angular indices which take values in $2\ldots 3$ we find
the non-zero components of the vector field covariant derivatives to be
\begin{align}
\nabla_0 A_0 =&  - e^{2(\Phi-\Psi)} A \Phi' \; ,
\\
\nabla_0 A_1 =& e^{\Phi} \chi \Phi' \; ,
\\
\nabla_1 A_0 =& -e^{\Phi} \chi' \; ,
\\
\nabla_1 A_1 =&  A'  - A \Psi'\; ,
\\
\nabla_A A_B =&   r e^{-2\Psi} A \Sigma_{AB}\;  .
\end{align}
where $\Sigma_{AB}$ is the metric tensor of a 2-sphere such that $d\Omega^2 = \Sigma_{AB} dx^A dx^B$.
The above relations lead to the definition of $E$ as
\begin{align}
E \equiv e^{-\Phi} F_{01} =  \chi' +  \chi \Phi'
\end{align}
while we get
\begin{align}
J_0 =&  -e^{\Phi-2\Psi} A E
\\
J_1 =&    \chi E
\\
J=&   e^{-2\Psi} \left[  A'  + A \left(\Phi' - \Psi' +  \frac{2}{r} \right)  \right]  
\end{align}
and
\begin{align}
F_{\mu\nu} F^{\mu\nu} =& -2e^{-2\Psi} E^2
\\
\nabla_\mu F^\mu_{\;\;0} =& - e^{\Phi-2\Psi}  \left[ E' + \left( - \Psi' + \frac{2}{r} \right) E \right]
\end{align}
while $\nabla_\mu F^\mu_{\;\;i} = 0$.

Useful components of Einstein and field equations in our coordinates include:
\begin{align}
q^\lambda_{\;\;\rho}\nabla_\lambda\nabla^\rho \phi = &
 \Qcal_0  e^{-2\Psi} \left[ 
\chi^2  R''
-  2 e^{-\Phi}  \chi A \qsol  \Phi'
+ \left( \frac{2}{r}  -  \chi^2  \Psi' -  e^{-2\Psi}  A^2   \Phi' \right)  R'
\right]
\\
J^\rho D_\rho \phi  =& \Qcal_0 e^{-2\Psi} E\left( e^{-\Phi} \qsol A  + \chi R'\right)
\\
\frac{C^{(g)}}{2-\KB} =&  \frac{\KB}{2-\KB} e^{-2\Psi} E^2
 + 2 \Qcal_0 e^{-2\Psi} E\left(e^{-\Phi} \qsol A  + \chi R'\right)
  \nonumber
\\
&
-  (1+\lambdas) \Qcal_0^2 e^{-2\Psi}  \left(e^{-\Phi} \qsol A  + \chi R'\right)^2
+ \mu^2 \left( C^{(\Qcal)} \right)^2 
\label{Cg_SSS}
\\
\frac{C^{(A)}}{2-\KB} =&  
 \frac{\KB }{2-\KB} e^{-2\Psi}  \chi\left[ E' + \left( - \Psi' + \frac{2}{r} \right) E \right]
 -  \Qcal_0 e^{-2\Psi} E\left(e^{-\Phi} \qsol A   + \chi R'\right)
 \nonumber
\\
&
+  \Qcal_0  e^{-2\Psi} \left[  A'  + A \left(\Phi' - \Psi' +  \frac{2}{r} \right)  \right] \left( e^{-\Phi}   \chi \qsol  +  e^{-2\Psi}A R' \right)
\nonumber
\\
&
+ \Qcal_0  e^{-2\Psi} \left[ \chi^2  R'' -  2 e^{-\Phi}  \chi A  \qsol  \Phi' + \left( \frac{2}{r}  -  \chi^2  \Psi' -  e^{-2\Psi}  A^2   \Phi' \right)  R' \right]
\nonumber
\\
&
+ \mu^2 C^{(\Qcal)} \left( e^{-\Phi}   \chi \qsol +  e^{-2\Psi}A R' \right)
\label{CA_SSS}
\end{align}

\section{Derivation of algebraically special solutions} 
\label{sec:appendix_A_special_q} 

\subsection{Case $\qsol=1$}
\label{app_A_0_q_0}
With $\qsol =1$, \eqref{scalar_charge} implies $\phi_0=0$ and so with $A=0$ \eqref{reduced_scalar_SSS} leads to
\begin{align}
(1+\lambdas)\Qcal_0 R' =    \chi E   =   \Phi'   .
\label{reduced_scalar_SSS_A_q1}
\end{align}
 The Einstein equation \eqref{G_01_SSS} leads to $\mu^2 R' $, which when combined with \eqref{reduced_scalar_SSS_A_q1} is consistent with the vector equation
\eqref{reduced_vector_SSS}. Thus either $R'=0$ or $\mu=0$. 

 In the case $R'=0$, \eqref{reduced_scalar_SSS_A_q1} leads to $\Phi'=0 = E$, and so we may set $\Phi=0$. Then the remaining of the Einstein equations give, in the case
 $\chi=1$, that $\Psi= 0$, and hence, the resulting spacetime is Minkowski. In the case $\chi=-1$, the Einstein equations give full consistency only iff 
 $\mu=0 = \Psi$, and thus the resulting spacetime is once more Minkowski.

Let us now move to the non-trivial case for which $R'\ne 0$, and thus by our considerations above, $\mu=0$. Then using \eqref{reduced_scalar_SSS_A_q1}  into
\eqref{Cg_SSS} and \eqref{CA_SSS} and then into the Einstein equations \eqref{reduced_Ein_00}, \eqref{reduced_Ein_11} and \eqref{reduced_Ein_AB}
leads, after some computation, to
\begin{align}
 e^{2\Psi}   &= 1 +   2r  \Phi'   + \nt r^2 (\Phi')^2,
 \label{Psi_equation_A_0}
\\
 \Phi'' &+ \left[ \frac{2}{r}  + 2 \Phi' +  \nt r \left(\Phi' \right)^2   \right] \Phi' =  0,
 \label{Phi_equation_A_0}
\end{align}
where $\nt$ is given by \eqref{def_nt}~\footnote{
After finishing this work, it was brought to our attention by William Barker, that AeST contains Eling-Jacobson solutions~\cite{Eling:2006df}.
Upon closer inspection, we checked that these are identical to our algebraically special branch, specifically,
 \eqref{Psi_equation_A_0} and \eqref{Phi_equation_A_0} are equivalent to (25) and (26) of~\cite{Eling:2006df} 
, as well as, (5.11) and (5.17) of \cite{Oost:2021tqi}, respectively. 
We thank W. Barker who brought this possible connection to our attention, and who has discovered these AeST solutions in an independent work 
with collaborators~\cite{Wormhole_TBS}.}.
To solve the above equations, we change variables to $x \equiv \ln (r/r_0)$ for some arbitrary scale $r_0$, and further define the variable
\begin{align}
u \equiv \frac{d\Phi}{dx}.
\label{u_def}
\end{align}
Thus, \eqref{Psi_equation_A_0} is rewritten as
\begin{align}
 e^{2\Psi}   &= 1 +   2 u   + \nt u^2 = \nt (u - u_1) (u - u_2) , 
 \label{Psi_equation_A_0_u}
 \end{align}
 where the second equality is possible because $0<\nt\leq 1$, so that the quadratic $1/\nt + 2 u /\nt  +    u^2 $ has two real roots $u_1$ and $u_2$. These two roots are both negative, that is,
\begin{align}
u_1 =& - \frac{1}{1+n},  
\qquad&
u_2 =& - \frac{1}{1-n}, 
\end{align}
with $ u_2 < u_1 < 0$,  where $n \equiv \sqrt{1 - \nt}$, such that, $ 0 < n < 1$. 

With this in mind, we write \eqref{Phi_equation_A_0} as  
\begin{align}
\frac{dx}{du}  &= - \frac{1}{ \nt (u - u_1) (u - u_2)  u} , 
 \label{Phi_equation_A_0_u}
\end{align}
which may  be split  using partial fractions, considering also that $u_1 - u_2  = 2n / \nt >0$, and then integrated to get
\begin{align}
r(u)  &=    \frac{r_0}{|u|} 
  \frac{|u-u_1|^{\frac{1+n}{2n}}}{|u-u_2|^{\frac{1-n}{2n}} },
\label{r_u_solution_App}
\end{align}
which is \eqref{r_u_solution}.
This, in fact means that there are four solution branches: $\{ u<u_2,  u_2 < u < u_1 , u_1<u<0, u>0 \}$ with $r=\infty$ corresponding to $u=0$ or to $u=u_2$.

Finally, we find $\Phi(u)$ by combining \eqref{u_def} and \eqref{Phi_equation_A_0_u} to get
\begin{align}
\Phi  =&  - \int \frac{du}{ \nt (u - u_1) (u - u_2)  } = \frac{1}{ 2n } \ln \frac{|u-u_2|}{|u-u_1|}. 
\label{Phi_of_u}
\end{align}
Thus with \eqref{Psi_equation_A_0_u}, \eqref{r_u_solution_App} and \eqref{Phi_of_u}, we reconstruct the metric \eqref{g_special_A_1} and so \eqref{phi_special_A_1}
also follows.

 \subsection{Case $\qsol=0$} 
In this case $\phi_0$ may take in principle any real value, and does not have to be zero. Then, with $A=\qsol=0$, which imply  
$ C^{(\Qcal)} = -1$, 
\eqref{reduced_scalar_SSS} gives
\begin{align}
   \Phi' - (1+\lambdas)\Qcal_0 R' = \frac{\phit_0  e^{-\Phi+\Psi} }{r^2},
\label{reduced_scalar_SSS_A_q0}
\end{align}
where $\phi_0 = \phit_0 \Qcal_0$, while both \eqref{reduced_vector_SSS} and \eqref{G_01_SSS}  lead to $\mu^2   R' =0$. 
Meanwhile 
the \eqref{reduced_Ein_00}  Einstein equations results in
\begin{align}
  \frac{2}{r}   \Psi'   +\frac{ e^{2\Psi}  - 1}{r^2}    =& 
\KB    \left[ \Phi'' 
+ \left( \frac{1}{2}  \Phi' - \Psi' + \frac{2}{r} \right) \Phi' 
\right]  
-   \frac{1}{2} (2-\KB) \mu^2e^{2\Psi}
  \nonumber
\\
& 
+  (2-\KB) \Qcal_0   \left[   R''  + \left( \frac{2}{r}  -    \Psi' + \frac{1}{2} (1+\lambdas) \Qcal_0   R' \right)  R'  \right],
\label{Ein_00_A_0_q_0}
\end{align}
\eqref{reduced_Ein_11} reduces to
\begin{align}
 \frac{2}{r}   \Phi'   -\frac{e^{2\Psi}-1}{r^2}   
  =& 
     - \frac{1}{2}  \KB   {\Phi'}^2 
 +(2-\KB) \left[ 
    -\Qcal_0   \Phi' R'  
    +\frac{1}{2}  (1+\lambdas) \Qcal_0^2  {R'}^2     
  +\frac{1}{2}  \mu^2 e^{2\Psi} \right], 
  \label{Ein_11_A_0_q_0}
\end{align}
and \eqref{reduced_Ein_AB} to
\begin{align}
\Phi'' + \left(\Phi' - \Psi' \right) \left(\Phi' +  \frac{1}{r} \right) 
 =&  
  (2-\KB)     \Qcal_0  R'   \left[ \Phi' - \frac{1}{2} (1+\lambdas) \Qcal_0     R' \right]
+ \frac{\KB  }{2} {\Phi'}^2 
\nonumber 
\\
& + \frac{1}{2} (2-\KB) \mu^2 e^{2\Psi}. 
\label{Ein_AB_A_0_q_0}
\end{align}

We now consider the two possible subcases: $R'=0$ and $\mu=0$ separately.

\subsubsection{Case $R'=0$}
Taking the derivative of  \eqref{reduced_scalar_SSS_A_q0}  gives
\begin{align}
 \Phi'' + (\Phi' - \Psi') \Phi' + \frac{2}{r} \Phi' = 0,
 \label{reduced_scalar_SSS_A_q0_der}
\end{align}
and adding \eqref{Ein_11_A_0_q_0} and \eqref{Ein_00_A_0_q_0} leads to
\begin{align}
   \frac{2}{r}  (\Phi'+ \Psi') =&  \KB    \left[ \Phi''  + \left(   - \Psi' + \frac{2}{r} \right) \Phi' \right]  .
   \label{added_Ein_00_11_A_q0}
\end{align}
 Then, using \eqref{reduced_scalar_SSS_A_q0_der} to 
eliminate $\Phi''$ allows us to express $\Psi'$ in terms of $\Phi'$ as
\begin{align}
  \Psi'   =&  - \frac{1}{2} \KB    r {\Phi'}^2   - \Phi',
  \label{Psi_Phi_A_0_q_0}
\end{align}
while \eqref{reduced_scalar_SSS_A_q0_der} subbed into  \eqref{Ein_AB_A_0_q_0} gives after using \eqref{reduced_scalar_SSS_A_q0_der}, 
the additional condition $\mu^2=0$.
 
Substituting into \eqref{Ein_00_A_0_q_0} leads to
\begin{align}
 e^{2\Psi} =& 1  + 2   r \Phi' + \frac{\KB}{2}   r^2 (\Phi')^2, 
 \label{exppsi_A_0_q_0}
\end{align}
and differentiating the above equation we eliminate $\Psi'$ with \eqref{Psi_Phi_A_0_q_0} to get
\begin{align}
\Phi'' +\frac{2}{r}   \Phi' +   2(\Phi')^2 +  \frac{\KB}{2} r      (\Phi')^3  =  0.
\label{phi_A_0_q_0}
\end{align}
The above equation admits a $\Phi' = \Psi' = 0$ Minkowski solution (obtainable also directly by $\phit_0=0$).

The system \eqref{exppsi_A_0_q_0} and \eqref{phi_A_0_q_0} is analogous to \eqref{Psi_equation_A_0} and \eqref{Phi_equation_A_0} with $\nt \rightarrow \KB/2$ whose solution we have already found. 
Thus, this system reduces to the $A=0, \qsol=1$ case studied in Subsection \ref{app_A_0_q_0}, that is, \eqref{Psi_equation_A_0_u} and \eqref{Phi_of_u}, with $r= r(u)$ given by \eqref{r_u_solution_App}. However, there are some additional constraints that make it different from the full solution of Subsection \ref{app_A_0_q_0},
 the first being that since $R'=0$ then $\phi = 0$ throughout. The second constraint that needs to be satisfied is in fact \eqref{reduced_scalar_SSS_A_q0}, that is,
\begin{align}
  \Phi'  = \frac{\phit_0  e^{-\Phi+\Psi} }{r^2},
\end{align}
and this leads to identifying  $\Qcal_0 r_0  =   |\phi_0|  \sqrt{\frac{\KB}{2}}$ and $\Sign{u} = \Sign{\phi_0}$. 

\subsubsection{Case $\mu=0$}
In this case adding \eqref{Ein_11_A_0_q_0} and \eqref{Ein_AB_A_0_q_0} we get
\begin{align}
\Phi'' + \left(\Phi' - \Psi' \right) \left(\Phi' +  \frac{1}{r} \right) 
+  \frac{2}{r}   \Phi'   +\frac{1-e^{2\Psi}}{r^2}   
 =&  
 0,
\label{eq_A_0_q_0_mu_0_Phipp}
\end{align}
while \eqref{reduced_scalar_SSS_A_q0} is used to solve for $R'$ as
\begin{align}
  (1+\lambdas)\Qcal_0 R' =    \Phi'-\frac{\phit_0  e^{-\Phi+\Psi} }{r^2}.
\label{Rp_q_0_mu_0_app}
\end{align}
We use the above equation to eliminate $R'$  so that  \eqref{Ein_00_A_0_q_0} leads to
\begin{align}
  \frac{2}{r}   \Psi'   +\frac{ e^{2\Psi}  - 1}{r^2}    =& 
   2 \nt      \left[ \Phi'' +   \left( \frac{2}{r}  -    \Psi' +  \frac{1}{2} \Phi'    \right)   \Phi'    \right]
+  \frac{2-\KB}{2(1+\lambdas)}   \frac{\phit_0^2  e^{-2\Phi+2\Psi} }{r^4}   ,
\end{align}
and \eqref{Ein_11_A_0_q_0} to
\begin{align}
 \frac{2}{r}   \Phi'   -\frac{e^{2\Psi}-1}{r^2}     =&   - \nt   {\Phi'}^2 + \frac{2-\KB}{2(1+\lambdas)}  \frac{\phit_0^2  e^{-2\Phi+2\Psi} }{r^4} ,
\label{G_00_q_0_mu_0_app}
\end{align}
so that subtracting one from the other to cancel out the  term proportional to $\phit_0^2$ we find
\begin{align}
  \frac{1}{r}  ( \Psi'  -\Phi')  + \frac{ e^{2\Psi}  - 1}{r^2}   =& 
    \nt      \left[ \Phi'' +   \left(   \Phi'   -    \Psi'  + \frac{2}{r}   \right)   \Phi'    \right].
\end{align}
Thus using \eqref{eq_A_0_q_0_mu_0_Phipp}  to eliminate the $\Phi''$ term  we find 
\begin{align} 
\Phi' = \Psi'    + \frac{ e^{2\Psi}  - 1}{r} ,
\end{align}
which allows us to eliminate $\Phi'$ from \eqref{eq_A_0_q_0_mu_0_Phipp} to get
\begin{align}
  \Psi'' 
+ \frac{1}{r} \left( 3e^{2\Psi}  + 1 \right)  \Psi' 
+ \frac{1}{r^2} e^{2\Psi} \left(  e^{2\Psi}  - 1 \right) 
  =&
0.
\label{eq_A_0_q_0_mu_0_Phippp}
\end{align}

\section{The $\qsol=0$ Schwarzschild case}
\label{app:q_0_Schwarzschild}
Taking the $\qsol=0$ solutions and setting $\charge=0$ leads to the Schwarzschild for the metric, while $\chi$ and $E$ in \eqref{chi_q_0_complete} reduce to
\begin{align}
\chi =&   e^{-\Phi} \sqrt{1 + \frac{2\qA}{r}  },
 \quad  & E =& - \frac{e^{-\Phi}}{r^2} \frac{   \qA  }{ \sqrt{1 + \frac{2 \qA}{r}   } }.
\label{E_chi_Phi_q0_q_0_Schwarzschild}
\end{align}
Likewise, \eqref{eq_A_q_0_sol} and \eqref{eq_phi_q_0_sol}  reduce respectively to
\begin{align}
   A =&  \sA e^{-2\Phi} \sqrt{  \frac{2 (\GN M +\qA)}{r}   },
\quad
& 
(1+\lambdas)\phi =& \ln \sqrt{  1  +\frac{2 \qA }{r} } +  \frac{ s_{\phi_0} }{2 \mt  } 
  \ln\frac{ 1 + \frac{\qA + \mt |\phit_0| }{ r}  }{ 1 + \frac{\qA - \mt |\phit_0|   }{ r }   },
   \label{eq_A_phi_q_0_sol_Schwarzschild}
\end{align}
while further setting $\phit_0 = 0$, leads to $\phi=0$, i.e. the scalar hair vanishes and we are left with only the vector hair.

\bibliographystyle{JHEP.bst}    
\bibliography{AeST_BH_article.bib}

\providecommand{\noopsort}[1]{}\providecommand{\singleletter}[1]{#1}%

\providecommand{\href}[2]{#2}\begingroup\raggedright\begin{thebibliography}{100}

\bibitem{will2014confrontation}
C.M.~Will, \emph{The confrontation between general relativity and experiment},
  {\emph{Living reviews in relativity} {\bfseries 17} (2014) 1}.

\bibitem{adelberger2003tests}
E.G.~Adelberger, B.R.~Heckel and A.E.~Nelson, \emph{Tests of the gravitational
  inverse-square law}, {\emph{arXiv preprint hep-ph/0307284} (2003) }.

\bibitem{ni2016solar}
W.-T.~Ni, \emph{Solar-system tests of the relativistic gravity},
  {\emph{International Journal of Modern Physics D} {\bfseries 25} (2016)
  1630003}.

\bibitem{berti2015testing}
E.~Berti, E.~Barausse, V.~Cardoso, L.~Gualtieri, P.~Pani, U.~Sperhake et~al.,
  \emph{Testing general relativity with present and future astrophysical
  observations}, {\emph{Classical and Quantum Gravity} {\bfseries 32} (2015)
  243001}.

\bibitem{PhysRevLett.119.161101}
{\scshape LIGO Scientific Collaboration and Virgo Collaboration} collaboration,
  \emph{Gw170817: Observation of gravitational waves from a binary neutron star
  inspiral}, \href{https://doi.org/10.1103/PhysRevLett.119.161101}{\emph{Phys.
  Rev. Lett.} {\bfseries 119} (2017) 161101}.

\bibitem{Clifton:2011jh}
T.~Clifton, P.G.~Ferreira, A.~Padilla and C.~Skordis, \emph{{Modified Gravity
  and Cosmology}},
  \href{https://doi.org/10.1016/j.physrep.2012.01.001}{\emph{Phys. Rept.}
  {\bfseries 513} (2012) 1} [\href{https://arxiv.org/abs/1106.2476}{{\ttfamily
  1106.2476}}].

\bibitem{Donoghue:1994dn}
J.F.~Donoghue, \emph{{General relativity as an effective field theory: The
  leading quantum corrections}},
  \href{https://doi.org/10.1103/PhysRevD.50.3874}{\emph{Phys. Rev. D}
  {\bfseries 50} (1994) 3874}
  [\href{https://arxiv.org/abs/gr-qc/9405057}{{\ttfamily gr-qc/9405057}}].

\bibitem{Burgess:2003jk}
C.P.~Burgess, \emph{{Quantum gravity in everyday life: General relativity as an
  effective field theory}},
  \href{https://doi.org/10.12942/lrr-2004-5}{\emph{Living Rev. Rel.} {\bfseries
  7} (2004) 5} [\href{https://arxiv.org/abs/gr-qc/0311082}{{\ttfamily
  gr-qc/0311082}}].

\bibitem{Woodard:2009ns}
R.P.~Woodard, \emph{{How Far Are We from the Quantum Theory of Gravity?}},
  \href{https://doi.org/10.1088/0034-4885/72/12/126002}{\emph{Rept. Prog.
  Phys.} {\bfseries 72} (2009) 126002}
  [\href{https://arxiv.org/abs/0907.4238}{{\ttfamily 0907.4238}}].

\bibitem{Joyce:2014kja}
A.~Joyce, B.~Jain, J.~Khoury and M.~Trodden, \emph{{Beyond the Cosmological
  Standard Model}},
  \href{https://doi.org/10.1016/j.physrep.2014.12.002}{\emph{Phys. Rept.}
  {\bfseries 568} (2015) 1} [\href{https://arxiv.org/abs/1407.0059}{{\ttfamily
  1407.0059}}].

\bibitem{Joyce:2016vqv}
A.~Joyce, L.~Lombriser and F.~Schmidt, \emph{{Dark Energy Versus Modified
  Gravity}},
  \href{https://doi.org/10.1146/annurev-nucl-102115-044553}{\emph{Ann. Rev.
  Nucl. Part. Sci.} {\bfseries 66} (2016) 95}
  [\href{https://arxiv.org/abs/1601.06133}{{\ttfamily 1601.06133}}].

\bibitem{Milgrom:1983ca}
M.~Milgrom, \emph{{A Modification of the Newtonian dynamics as a possible
  alternative to the hidden mass hypothesis}},
  \href{https://doi.org/10.1086/161130}{\emph{Astrophys. J.} {\bfseries 270}
  (1983) 365}.

\bibitem{Milgrom:1983pn}
M.~Milgrom, \emph{{A Modification of the Newtonian dynamics: Implications for
  galaxies}}, \href{https://doi.org/10.1086/161131}{\emph{Astrophys. J.}
  {\bfseries 270} (1983) 371}.

\bibitem{Milgrom:1983zz}
M.~Milgrom, \emph{{A modification of the Newtonian dynamics: implications for
  galaxy systems}}, \href{https://doi.org/10.1086/161132}{\emph{Astrophys. J.}
  {\bfseries 270} (1983) 384}.

\bibitem{famaey2012modified}
B.~Famaey and S.S.~McGaugh, \emph{Modified newtonian dynamics (mond):
  observational phenomenology and relativistic extensions}, {\emph{Living
  reviews in relativity} {\bfseries 15} (2012) 1}.

\bibitem{1984ApJ...286....7B}
J.~{Bekenstein} and M.~{Milgrom}, \emph{{Does the missing mass problem signal
  the breakdown of Newtonian gravity?}},
  \href{https://doi.org/10.1086/162570}{\emph{Astrophys. J.} {\bfseries 286}
  (1984) 7}.

\bibitem{Bekenstein:1988zy}
J.D.~Bekenstein, \emph{{Phase Coupling Gravitation: Symmetries and Gauge
  Fields}}, \href{https://doi.org/10.1016/0370-2693(88)91851-5}{\emph{Phys.
  Lett. B} {\bfseries 202} (1988) 497}.

\bibitem{Bekenstein:1992pj}
J.D.~Bekenstein, \emph{{The Relation between physical and gravitational
  geometry}}, \href{https://doi.org/10.1103/PhysRevD.48.3641}{\emph{Phys. Rev.
  D} {\bfseries 48} (1993) 3641}
  [\href{https://arxiv.org/abs/gr-qc/9211017}{{\ttfamily gr-qc/9211017}}].

\bibitem{Bekenstein:1993fs}
J.D.~Bekenstein and R.H.~Sanders, \emph{{Gravitational lenses and
  unconventional gravity theories}},
  \href{https://doi.org/10.1086/174337}{\emph{Astrophys. J.} {\bfseries 429}
  (1994) 480} [\href{https://arxiv.org/abs/astro-ph/9311062}{{\ttfamily
  astro-ph/9311062}}].

\bibitem{Sanders:1996wk}
R.H.~Sanders, \emph{{A Stratified framework for scalar - tensor theories of
  modified dynamics}}, \href{https://doi.org/10.1086/303980}{\emph{Astrophys.
  J.} {\bfseries 480} (1997) 492}
  [\href{https://arxiv.org/abs/astro-ph/9612099}{{\ttfamily
  astro-ph/9612099}}].

\bibitem{PhysRevD.70.083509}
J.D.~Bekenstein, \emph{Relativistic gravitation theory for the modified
  newtonian dynamics paradigm},
  \href{https://doi.org/10.1103/PhysRevD.70.083509}{\emph{Phys. Rev. D}
  {\bfseries 70} (2004) 083509}.

\bibitem{Skordis:2005xk}
C.~Skordis, D.F.~Mota, P.G.~Ferreira and C.~Boehm, \emph{{Large Scale Structure
  in Bekenstein's theory of relativistic Modified Newtonian Dynamics}},
  \href{https://doi.org/10.1103/PhysRevLett.96.011301}{\emph{Phys. Rev. Lett.}
  {\bfseries 96} (2006) 011301}
  [\href{https://arxiv.org/abs/astro-ph/0505519}{{\ttfamily
  astro-ph/0505519}}].

\bibitem{Dodelson:2006zt}
S.~Dodelson and M.~Liguori, \emph{{Can Cosmic Structure form without Dark
  Matter?}}, \href{https://doi.org/10.1103/PhysRevLett.97.231301}{\emph{Phys.
  Rev. Lett.} {\bfseries 97} (2006) 231301}
  [\href{https://arxiv.org/abs/astro-ph/0608602}{{\ttfamily
  astro-ph/0608602}}].

\bibitem{Bourliot:2006ig}
F.~Bourliot, P.G.~Ferreira, D.F.~Mota and C.~Skordis, \emph{{The cosmological
  behavior of Bekenstein's modified theory of gravity}},
  \href{https://doi.org/10.1103/PhysRevD.75.063508}{\emph{Phys. Rev. D}
  {\bfseries 75} (2007) 063508}
  [\href{https://arxiv.org/abs/astro-ph/0611255}{{\ttfamily
  astro-ph/0611255}}].

\bibitem{giannios2005spherically}
D.~Giannios, \emph{Spherically symmetric, static spacetimes in a
  tensor-vector-scalar theory}, {\emph{Physical Review D} {\bfseries 71} (2005)
  103511}.

\bibitem{sagi2008black}
E.~Sagi and J.D.~Bekenstein, \emph{Black holes in the tensor-vector-scalar
  theory of gravity and their thermodynamics}, {\emph{Physical Review D}
  {\bfseries 77} (2008) 024010}.

\bibitem{Lasky:2010bd}
P.D.~Lasky and D.D.~Doneva, \emph{{Stability and Quasinormal Modes of Black
  holes in Tensor-Vector-Scalar theory: Scalar Field Perturbations}},
  \href{https://doi.org/10.1103/PhysRevD.82.124068}{\emph{Phys. Rev. D}
  {\bfseries 82} (2010) 124068}
  [\href{https://arxiv.org/abs/1011.0747}{{\ttfamily 1011.0747}}].

\bibitem{lasky2008structure}
P.D.~Lasky, H.~Sotani and D.~Giannios, \emph{Structure of neutron stars in
  tensor-vector-scalar theory}, {\emph{Physical Review D} {\bfseries 78} (2008)
  104019}.

\bibitem{Sagi:2009kd}
E.~Sagi, \emph{{Preferred frame parameters in the tensor-vector-scalar theory
  of gravity and its generalization}},
  \href{https://doi.org/10.1103/PhysRevD.80.044032}{\emph{Phys. Rev. D}
  {\bfseries 80} (2009) 044032}
  [\href{https://arxiv.org/abs/0905.4001}{{\ttfamily 0905.4001}}].

\bibitem{Sagi:2010ei}
E.~Sagi, \emph{{Propagation of gravitational waves in generalized TeVeS}},
  \href{https://doi.org/10.1103/PhysRevD.81.064031}{\emph{Phys. Rev. D}
  {\bfseries 81} (2010) 064031}
  [\href{https://arxiv.org/abs/1001.1555}{{\ttfamily 1001.1555}}].

\bibitem{Chaichian:2014dfa}
M.~Chaichian, J.~Kluso\v{n}, M.~Oksanen and A.~Tureanu, \emph{{Can TeVeS be a
  viable theory of gravity?}},
  \href{https://doi.org/10.1016/j.physletb.2014.06.036}{\emph{Phys. Lett. B}
  {\bfseries 735} (2014) 322}
  [\href{https://arxiv.org/abs/1402.4696}{{\ttfamily 1402.4696}}].

\bibitem{Gong:2018cgj}
Y.~Gong, S.~Hou, D.~Liang and E.~Papantonopoulos, \emph{{Gravitational waves in
  Einstein-\ae{}ther and generalized TeVeS theory after GW170817}},
  \href{https://doi.org/10.1103/PhysRevD.97.084040}{\emph{Phys. Rev. D}
  {\bfseries 97} (2018) 084040}
  [\href{https://arxiv.org/abs/1801.03382}{{\ttfamily 1801.03382}}].

\bibitem{Skordis:2019fxt}
C.~Skordis and T.~Z\l{}o\'snik, \emph{{Gravitational alternatives to dark
  matter with tensor mode speed equaling the speed of light}},
  \href{https://doi.org/10.1103/PhysRevD.100.104013}{\emph{Phys. Rev. D}
  {\bfseries 100} (2019) 104013}
  [\href{https://arxiv.org/abs/1905.09465}{{\ttfamily 1905.09465}}].

\bibitem{Sanders:2005vd}
R.H.~Sanders, \emph{{A Tensor-vector-scalar framework for modified dynamics and
  cosmic dark matter}},
  \href{https://doi.org/10.1111/j.1365-2966.2005.09375.x}{\emph{Mon. Not. Roy.
  Astron. Soc.} {\bfseries 363} (2005) 459}
  [\href{https://arxiv.org/abs/astro-ph/0502222}{{\ttfamily
  astro-ph/0502222}}].

\bibitem{Skordis:2008pq}
C.~Skordis, \emph{{Generalizing tensor-vector-scalar cosmology}},
  \href{https://doi.org/10.1103/PhysRevD.77.123502}{\emph{Phys. Rev. D}
  {\bfseries 77} (2008) 123502}
  [\href{https://arxiv.org/abs/0801.1985}{{\ttfamily 0801.1985}}].

\bibitem{babichev2011improving}
E.~Babichev, C.~Deffayet and G.~Esposito-Farese, \emph{Improving relativistic
  mond with galileon k-mouflage https://doi. org/10.1103/physrevd. 84.061502
  phys. rev}, {\emph{D} {\bfseries 84} (2011) 1106}.

\bibitem{Zlosnik:2017xpr}
T.G.~Z\l{}o\'snik and C.~Skordis, \emph{{Cosmology of the Galileon extension of
  Bekenstein\textquoteright{}s theory of relativistic modified Newtonian
  dynamics}}, \href{https://doi.org/10.1103/PhysRevD.95.124023}{\emph{Phys.
  Rev. D} {\bfseries 95} (2017) 124023}
  [\href{https://arxiv.org/abs/1702.00683}{{\ttfamily 1702.00683}}].

\bibitem{Zlosnik:2006zu}
T.G.~Zlosnik, P.G.~Ferreira and G.D.~Starkman, \emph{{Modifying gravity with
  the Aether: An alternative to Dark Matter}},
  \href{https://doi.org/10.1103/PhysRevD.75.044017}{\emph{Phys. Rev. D}
  {\bfseries 75} (2007) 044017}
  [\href{https://arxiv.org/abs/astro-ph/0607411}{{\ttfamily
  astro-ph/0607411}}].

\bibitem{PhysRevD.80.123536}
M.~Milgrom, \emph{Bimetric mond gravity},
  \href{https://doi.org/10.1103/PhysRevD.80.123536}{\emph{Phys. Rev. D}
  {\bfseries 80} (2009) 123536}.

\bibitem{Zuntz:2010jp}
J.~Zuntz, T.G.~Zlosnik, F.~Bourliot, P.G.~Ferreira and G.D.~Starkman,
  \emph{{Vector field models of modified gravity and the dark sector}},
  \href{https://doi.org/10.1103/PhysRevD.81.104015}{\emph{Phys. Rev. D}
  {\bfseries 81} (2010) 104015}
  [\href{https://arxiv.org/abs/1002.0849}{{\ttfamily 1002.0849}}].

\bibitem{Blanchet:2011wv}
L.~Blanchet and S.~Marsat, \emph{{Modified gravity approach based on a
  preferred time foliation}},
  \href{https://doi.org/10.1103/PhysRevD.84.044056}{\emph{Phys. Rev. D}
  {\bfseries 84} (2011) 044056}
  [\href{https://arxiv.org/abs/1107.5264}{{\ttfamily 1107.5264}}].

\bibitem{Sanders:2011wa}
R.H.~Sanders, \emph{{Hiding Lorentz Invariance Violation with MOND}},
  \href{https://doi.org/10.1103/PhysRevD.84.084024}{\emph{Phys. Rev. D}
  {\bfseries 84} (2011) 084024}
  [\href{https://arxiv.org/abs/1105.3910}{{\ttfamily 1105.3910}}].

\bibitem{deffayet2011nonlocal}
C.~Deffayet, G.~Esposito-Farese and R.P.~Woodard, \emph{Nonlocal metric
  formulations of modified newtonian dynamics with sufficient lensing},
  {\emph{Physical Review D—Particles, Fields, Gravitation, and Cosmology}
  {\bfseries 84} (2011) 124054}.

\bibitem{Mendoza:2012hu}
S.~Mendoza, T.~Bernal, J.C.~Hidalgo and S.~Capozziello, \emph{{MOND as the
  weak-field limit of an extended metric theory of gravity}},
  \href{https://doi.org/10.1063/1.4734465}{\emph{AIP Conf. Proc.} {\bfseries
  1458} (2012) 483} [\href{https://arxiv.org/abs/1202.3629}{{\ttfamily
  1202.3629}}].

\bibitem{Khoury:2014tka}
J.~Khoury, \emph{{Alternative to particle dark matter}},
  \href{https://doi.org/10.1103/PhysRevD.91.024022}{\emph{Phys. Rev. D}
  {\bfseries 91} (2015) 024022}
  [\href{https://arxiv.org/abs/1409.0012}{{\ttfamily 1409.0012}}].

\bibitem{deffayet2014field}
C.~Deffayet, G.~Esposito-Far{\`e}se and R.P.~Woodard, \emph{Field equations and
  cosmology for a class of nonlocal metric models of mond}, {\emph{Physical
  Review D} {\bfseries 90} (2014) 064038}.

\bibitem{Verlinde:2016toy}
E.P.~Verlinde, \emph{{Emergent Gravity and the Dark Universe}},
  \href{https://doi.org/10.21468/SciPostPhys.2.3.016}{\emph{SciPost Phys.}
  {\bfseries 2} (2017) 016} [\href{https://arxiv.org/abs/1611.02269}{{\ttfamily
  1611.02269}}].

\bibitem{Burrage:2018zuj}
C.~Burrage, E.J.~Copeland, C.~K\"ading and P.~Millington, \emph{{Symmetron
  scalar fields: Modified gravity, dark matter, or both?}},
  \href{https://doi.org/10.1103/PhysRevD.99.043539}{\emph{Phys. Rev. D}
  {\bfseries 99} (2019) 043539}
  [\href{https://arxiv.org/abs/1811.12301}{{\ttfamily 1811.12301}}].

\bibitem{PhysRevD.100.084039}
M.~Milgrom, \emph{Noncovariance at low accelerations as a route to mond},
  \href{https://doi.org/10.1103/PhysRevD.100.084039}{\emph{Phys. Rev. D}
  {\bfseries 100} (2019) 084039}.

\bibitem{DAmbrosio:2020nev}
F.~D'Ambrosio, M.~Garg and L.~Heisenberg, \emph{{Non-linear extension of
  non-metricity scalar for MOND}},
  \href{https://doi.org/10.1016/j.physletb.2020.135970}{\emph{Phys. Lett. B}
  {\bfseries 811} (2020) 135970}
  [\href{https://arxiv.org/abs/2004.00888}{{\ttfamily 2004.00888}}].

\bibitem{deffayet2024price}
C.~Deffayet and R.~Woodard, \emph{The price of abandoning dark matter is
  nonlocality}, {\emph{Journal of Cosmology and Astroparticle Physics}
  {\bfseries 2024} (2024) 042}.

\bibitem{Blanchet:2024mvy}
L.~Blanchet and C.~Skordis, \emph{{Relativistic Khronon theory in agreement
  with modified Newtonian dynamics and large-scale cosmology}},
  \href{https://doi.org/10.1088/1475-7516/2024/11/040}{\emph{JCAP} {\bfseries
  11} (2024) 040} [\href{https://arxiv.org/abs/2404.06584}{{\ttfamily
  2404.06584}}].

\bibitem{bruneton2007field}
J.-P.~Bruneton and G.~Esposito-Farese, \emph{Field-theoretical formulations of
  mond-like gravity}, {\emph{Physical Review D} {\bfseries 76} (2007) 124012}.

\bibitem{Blanchet:2006yt}
L.~Blanchet, \emph{{Gravitational polarization and the phenomenology of MOND}},
  \href{https://doi.org/10.1088/0264-9381/24/14/001}{\emph{Class. Quant. Grav.}
  {\bfseries 24} (2007) 3529}
  [\href{https://arxiv.org/abs/astro-ph/0605637}{{\ttfamily
  astro-ph/0605637}}].

\bibitem{Blanchet:2009zu}
L.~Blanchet and A.~Le~Tiec, \emph{{Dipolar Dark Matter and Dark Energy}},
  \href{https://doi.org/10.1103/PhysRevD.80.023524}{\emph{Phys. Rev. D}
  {\bfseries 80} (2009) 023524}
  [\href{https://arxiv.org/abs/0901.3114}{{\ttfamily 0901.3114}}].

\bibitem{Berezhiani:2015pia}
L.~Berezhiani and J.~Khoury, \emph{{Dark Matter Superfluidity and Galactic
  Dynamics}}, \href{https://doi.org/10.1016/j.physletb.2015.12.054}{\emph{Phys.
  Lett. B} {\bfseries 753} (2016) 639}
  [\href{https://arxiv.org/abs/1506.07877}{{\ttfamily 1506.07877}}].

\bibitem{Berezhiani:2015bqa}
L.~Berezhiani and J.~Khoury, \emph{{Theory of dark matter superfluidity}},
  \href{https://doi.org/10.1103/PhysRevD.92.103510}{\emph{Phys. Rev. D}
  {\bfseries 92} (2015) 103510}
  [\href{https://arxiv.org/abs/1507.01019}{{\ttfamily 1507.01019}}].

\bibitem{Kaplinghat:2015aga}
M.~Kaplinghat, S.~Tulin and H.-B.~Yu, \emph{{Dark Matter Halos as Particle
  Colliders: Unified Solution to Small-Scale Structure Puzzles from Dwarfs to
  Clusters}}, \href{https://doi.org/10.1103/PhysRevLett.116.041302}{\emph{Phys.
  Rev. Lett.} {\bfseries 116} (2016) 041302}
  [\href{https://arxiv.org/abs/1508.03339}{{\ttfamily 1508.03339}}].

\bibitem{Kamada:2016euw}
A.~Kamada, M.~Kaplinghat, A.B.~Pace and H.-B.~Yu, \emph{{How the
  Self-Interacting Dark Matter Model Explains the Diverse Galactic Rotation
  Curves}}, \href{https://doi.org/10.1103/PhysRevLett.119.111102}{\emph{Phys.
  Rev. Lett.} {\bfseries 119} (2017) 111102}
  [\href{https://arxiv.org/abs/1611.02716}{{\ttfamily 1611.02716}}].

\bibitem{Blanchet:2015sra}
L.~Blanchet and L.~Heisenberg, \emph{{Dark Matter via Massive (bi-)Gravity}},
  \href{https://doi.org/10.1103/PhysRevD.91.103518}{\emph{Phys. Rev. D}
  {\bfseries 91} (2015) 103518}
  [\href{https://arxiv.org/abs/1504.00870}{{\ttfamily 1504.00870}}].

\bibitem{LIGOScientific:2017vwq}
{\scshape LIGO Scientific, Virgo} collaboration, \emph{{GW170817: Observation
  of Gravitational Waves from a Binary Neutron Star Inspiral}},
  \href{https://doi.org/10.1103/PhysRevLett.119.161101}{\emph{Phys. Rev. Lett.}
  {\bfseries 119} (2017) 161101}
  [\href{https://arxiv.org/abs/1710.05832}{{\ttfamily 1710.05832}}].

\bibitem{Savchenko:2017ffs}
V.~Savchenko et~al., \emph{{INTEGRAL Detection of the First Prompt Gamma-Ray
  Signal Coincident with the Gravitational-wave Event GW170817}},
  \href{https://doi.org/10.3847/2041-8213/aa8f94}{\emph{Astrophys. J. Lett.}
  {\bfseries 848} (2017) L15}
  [\href{https://arxiv.org/abs/1710.05449}{{\ttfamily 1710.05449}}].

\bibitem{Goldstein:2017mmi}
A.~Goldstein et~al., \emph{{An Ordinary Short Gamma-Ray Burst with
  Extraordinary Implications: Fermi-GBM Detection of GRB 170817A}},
  \href{https://doi.org/10.3847/2041-8213/aa8f41}{\emph{Astrophys. J. Lett.}
  {\bfseries 848} (2017) L14}
  [\href{https://arxiv.org/abs/1710.05446}{{\ttfamily 1710.05446}}].

\bibitem{2021PhRvL.127p1302S}
C.~{Skordis} and T.~{Z{\l}o{\'s}nik}, \emph{{New Relativistic Theory for
  Modified Newtonian Dynamics}},
  \href{https://doi.org/10.1103/PhysRevLett.127.161302}{\emph{Phys. Rev. Lett.}
  {\bfseries 127} (2021) 161302}
  [\href{https://arxiv.org/abs/2007.00082}{{\ttfamily 2007.00082}}].

\bibitem{2022PhRvD.106j4041S}
C.~{Skordis} and T.~{Zlosnik}, \emph{{Aether scalar tensor theory: Linear
  stability on Minkowski space}},
  \href{https://doi.org/10.1103/PhysRevD.106.104041}{\emph{Phys. Rev. D}
  {\bfseries 106} (2022) 104041}
  [\href{https://arxiv.org/abs/2109.13287}{{\ttfamily 2109.13287}}].

\bibitem{Kashfi:2022dyb}
T.~Kashfi and M.~Roshan, \emph{{Cosmological dynamics of relativistic MOND}},
  \href{https://doi.org/10.1088/1475-7516/2022/10/029}{\emph{JCAP} {\bfseries
  10} (2022) 029} [\href{https://arxiv.org/abs/2204.05672}{{\ttfamily
  2204.05672}}].

\bibitem{Mistele:2021qvz}
T.~Mistele, \emph{{Cherenkov radiation from stars constrains hybrid MOND dark
  matter models}},
  \href{https://doi.org/10.1088/1475-7516/2022/11/008}{\emph{JCAP} {\bfseries
  11} (2022) 008} [\href{https://arxiv.org/abs/2103.16954}{{\ttfamily
  2103.16954}}].

\bibitem{PhysRevD.107.044062}
S.~Tian, S.~Hou, S.~Cao and Z.-H.~Zhu, \emph{Time evolution of the local
  gravitational parameters and gravitational wave polarizations in a
  relativistic mond theory},
  \href{https://doi.org/10.1103/PhysRevD.107.044062}{\emph{Phys. Rev. D}
  {\bfseries 107} (2023) 044062}.

\bibitem{2023A&A...676A.100M}
T.~{Mistele}, S.~{McGaugh} and S.~{Hossenfelder}, \emph{{Aether scalar tensor
  theory confronted with weak lensing data at small accelerations}},
  \href{https://doi.org/10.1051/0004-6361/202346025}{\emph{Astronomy and
  Astrophysics} {\bfseries 676} (2023) A100}
  [\href{https://arxiv.org/abs/2301.03499}{{\ttfamily 2301.03499}}].

\bibitem{2023GReGr..55...23B}
R.C.~{Bernardo} and C.-Y.~{Chen}, \emph{{Dressed black holes in the new
  tensor-vector-scalar theory}},
  \href{https://doi.org/10.1007/s10714-023-03075-x}{\emph{General Relativity
  and Gravitation} {\bfseries 55} (2023) 23}
  [\href{https://arxiv.org/abs/2202.08460}{{\ttfamily 2202.08460}}].

\bibitem{llinares2023extension}
C.~Llinares, \emph{Extension of general relativity with mond limit predicts
  novel orbital structure in and around galaxies}, {\emph{arXiv preprint
  arXiv:2302.12032} (2023) }.

\bibitem{Verwayen:2023sds}
P.~Verwayen, C.~Skordis and C.~B\oe{}hm, \emph{{Aether Scalar Tensor (AeST)
  theory: quasistatic spherical solutions and their phenomenology}},
  \href{https://doi.org/10.1093/mnras/stae1225}{\emph{Mon. Not. Roy. Astron.
  Soc.} {\bfseries 531} (2024) 272}
  [\href{https://arxiv.org/abs/2304.05134}{{\ttfamily 2304.05134}}].

\bibitem{Durakovic:2023out}
A.~Durakovic and C.~Skordis, \emph{{Towards galaxy cluster models in
  Aether-Scalar-Tensor theory: isothermal spheres and curiosities}},
  \href{https://doi.org/10.1088/1475-7516/2024/04/040}{\emph{JCAP} {\bfseries
  04} (2024) 040} [\href{https://arxiv.org/abs/2312.00889}{{\ttfamily
  2312.00889}}].

\bibitem{Bataki:2023uuy}
M.~Bataki, C.~Skordis and T.~Zlosnik, \emph{{Aether scalar-tensor theory:
  Hamiltonian formalism}},
  \href{https://doi.org/10.1103/PhysRevD.110.044015}{\emph{Phys. Rev. D}
  {\bfseries 110} (2024) 044015}
  [\href{https://arxiv.org/abs/2307.15126}{{\ttfamily 2307.15126}}].

\bibitem{Mistele:2023fwd}
T.~Mistele, \emph{{New scale in the quasi-static limit of aether scalar tensor
  theory}}, \href{https://doi.org/10.1103/PhysRevD.110.024062}{\emph{Phys. Rev.
  D} {\bfseries 110} (2024) 024062}
  [\href{https://arxiv.org/abs/2305.07742}{{\ttfamily 2305.07742}}].

\bibitem{Rosa:2023qun}
J.a.L.~Rosa and T.~Zlosnik, \emph{{Dynamical system analysis of cosmological
  evolution in the Aether scalar tensor theory}},
  \href{https://doi.org/10.1103/PhysRevD.109.024018}{\emph{Phys. Rev. D}
  {\bfseries 109} (2024) 024018}
  [\href{https://arxiv.org/abs/2309.06232}{{\ttfamily 2309.06232}}].

\bibitem{reyes2024neutronstarsaetherscalartensor}
C.~Reyes and J.~Sakstein, \emph{Neutron stars in aether scalar-tensor theory},
  2024.

\bibitem{babichev2015charged}
E.~Babichev, C.~Charmousis and M.~Hassaine, \emph{Charged galileon black
  holes}, {\emph{Journal of Cosmology and Astroparticle Physics} {\bfseries
  2015} (2015) 031}.

\bibitem{collaboration2019first}
E.H.T.~Collaboration, K.~Akiyama, A.~Alberdi, W.~Alef, K.~Asada, R.~AZULY
  et~al., \emph{First m87 event horizon telescope results. i. the shadow of the
  supermassive black hole}, {\emph{Astrophys. J. Lett} {\bfseries 875} (2019)
  L1}.

\bibitem{yagi2016black}
K.~Yagi and L.C.~Stein, \emph{Black hole based tests of general relativity},
  {\emph{Classical and Quantum Gravity} {\bfseries 33} (2016) 054001}.

\bibitem{coleman1992quantum}
S.~Coleman, J.~Preskill and F.~Wilczek, \emph{Quantum hair on black holes},
  {\emph{Nuclear Physics B} {\bfseries 378} (1992) 175}.

\bibitem{ayon2006stealth}
E.~Ayon-Beato, C.~Martinez and J.~Zanelli, \emph{Stealth scalar field
  overflying a 2+ 1 black hole}, {\emph{General Relativity and Gravitation}
  {\bfseries 38} (2006) 145}.

\bibitem{bakopoulos2024black}
A.~Bakopoulos, C.~Charmousis, P.~Kanti, N.~Lecoeur and T.~Nakas, \emph{Black
  holes with primary scalar hair}, {\emph{Physical Review D} {\bfseries 109}
  (2024) 024032}.

\bibitem{fan2018black}
Z.-Y.~Fan, \emph{Black holes in vector-tensor theories and their
  thermodynamics}, {\emph{The European Physical Journal C} {\bfseries 78}
  (2018) 1}.

\bibitem{baake2024endowing}
O.~Baake, A.~Cisterna, M.~Hassaine and U.~Hernandez-Vera, \emph{Endowing black
  holes with beyond-horndeski primary hair: An exact solution framework for
  scalarizing in every dimension}, {\emph{Physical Review D} {\bfseries 109}
  (2024) 064024}.

\bibitem{chagoya2018stealth}
J.~Chagoya and G.~Tasinato, \emph{Stealth configurations in vector-tensor
  theories of gravity}, {\emph{Journal of Cosmology and Astroparticle Physics}
  {\bfseries 2018} (2018) 046}.

\bibitem{de2019perturbations}
C.~de~Rham and J.~Zhang, \emph{Perturbations of stealth black holes in
  degenerate higher-order scalar-tensor theories}, {\emph{Physical Review D}
  {\bfseries 100} (2019) 124023}.

\bibitem{de2023approximately}
A.~De~Felice, S.~Mukohyama and K.~Takahashi, \emph{Approximately stealth black
  hole in higher-order scalar-tensor theories}, {\emph{Journal of Cosmology and
  Astroparticle Physics} {\bfseries 2023} (2023) 050}.

\bibitem{bakopoulos2024stealth}
A.~Bakopoulos, T.~Karakasis and E.~Papantonopoulos, \emph{To stealth or not to
  stealth: Thermodynamics of stealth black holes}, {\emph{arXiv preprint
  arXiv:2410.14451} (2024) }.

\bibitem{Erices:2024iah}
C.~Erices, L.~Guajardo and K.~Lara, \emph{{Reverse stealth construction and its
  thermodynamic imprints}},  \href{https://arxiv.org/abs/2410.13719}{{\ttfamily
  2410.13719}}.

\bibitem{minamitsuji2018stealth}
M.~Minamitsuji and H.~Motohashi, \emph{Stealth schwarzschild solution in shift
  symmetry breaking theories}, {\emph{Physical Review D} {\bfseries 98} (2018)
  084027}.

\bibitem{minamitsuji2019black}
M.~Minamitsuji and J.~Edholm, \emph{Black hole solutions in shift-symmetric
  degenerate higher-order scalar-tensor theories}, {\emph{Physical Review D}
  {\bfseries 100} (2019) 044053}.

\bibitem{bernardo2020stealth}
R.C.~Bernardo, J.~Celestial and I.~Vega, \emph{Stealth black holes in shift
  symmetric kinetic gravity braiding}, {\emph{Physical Review D} {\bfseries
  101} (2020) 024036}.

\bibitem{heisenberg2017hairy}
L.~Heisenberg, R.~Kase, M.~Minamitsuji and S.~Tsujikawa, \emph{Hairy black-hole
  solutions in generalized proca theories}, {\emph{Physical Review D}
  {\bfseries 96} (2017) 084049}.

\bibitem{minamitsuji2016solutions}
M.~Minamitsuji, \emph{Solutions in the generalized proca theory with the
  nonminimal coupling to the einstein tensor}, {\emph{Physical Review D}
  {\bfseries 94} (2016) 084039}.

\bibitem{Skordis:2009bf}
C.~Skordis, \emph{{The Tensor-Vector-Scalar theory and its cosmology}},
  \href{https://doi.org/10.1088/0264-9381/26/14/143001}{\emph{Class. Quant.
  Grav.} {\bfseries 26} (2009) 143001}
  [\href{https://arxiv.org/abs/0903.3602}{{\ttfamily 0903.3602}}].

\bibitem{wald:1984}
R.M.~Wald, \emph{{General Relativity}}, The University of Chicago Press (1984).

\bibitem{babichev2014dressing}
E.~Babichev and C.~Charmousis, \emph{Dressing a black hole with a
  time-dependent galileon}, {\emph{Journal of High Energy Physics} {\bfseries
  2014} (2014) 1}.

\bibitem{kobayashi2014exact}
T.~Kobayashi and N.~Tanahashi, \emph{Exact black hole solutions in shift
  symmetric scalar--tensor theories}, {\emph{Progress of Theoretical and
  Experimental Physics} {\bfseries 2014} (2014) 073E02}.

\bibitem{charmousis2015self}
C.~Charmousis and D.~Iosifidis, \emph{Self tuning scalar tensor black holes},
  in \emph{Journal of Physics: Conference Series}, vol.~600, p.~012003, IOP
  Publishing, 2015.

\bibitem{Eling:2006df}
C.~Eling and T.~Jacobson, \emph{{Spherical solutions in Einstein-aether theory:
  Static aether and stars}},
  \href{https://doi.org/10.1088/0264-9381/23/18/008}{\emph{Class. Quant. Grav.}
  {\bfseries 23} (2006) 5625}
  [\href{https://arxiv.org/abs/gr-qc/0603058}{{\ttfamily gr-qc/0603058}}].

\bibitem{Oost:2021tqi}
J.~Oost, S.~Mukohyama and A.~Wang, \emph{{Spherically Symmetric Exact Vacuum
  Solutions in Einstein-Aether Theory}},
  \href{https://doi.org/10.3390/universe7080272}{\emph{Universe} {\bfseries 7}
  (2021) 272} [\href{https://arxiv.org/abs/2106.09044}{{\ttfamily
  2106.09044}}].

\bibitem{Wormhole_TBS}
L.~Yang, W.~Barker, A.~Durakovic and T.~Mistele, \emph{Spherical wormholes and
  cosmology in einstein-aether and aether-scalar-tensor theory (tentative)},
  {\emph{(to be submitted)} }.

\bibitem{creminelli2020hairy}
P.~Creminelli, N.~Loayza, F.~Serra, E.~Trincherini and L.G.~Trombetta,
  \emph{{Hairy Black-holes in Shift-symmetric Theories}},
  \href{https://doi.org/10.1007/JHEP08(2020)045}{\emph{JHEP} {\bfseries 08}
  (2020) 045} [\href{https://arxiv.org/abs/2004.02893}{{\ttfamily
  2004.02893}}].

\bibitem{Babichev:2024txe}
E.~Babichev, I.~Sawicki and L.G.~Trombetta, \emph{{The cosmic trimmer:
  Black-hole hair in scalar-Gauss-Bonnet gravity is altered by cosmology}},
  \href{https://arxiv.org/abs/2403.15537}{{\ttfamily 2403.15537}}.

\end{thebibliography}\endgroup

\end{document}